\documentclass[10pt,a4paper]{article}
\usepackage{f1000_styles}

\usepackage[round]{natbib}

\usepackage{adjustbox}
\usepackage{xspace}
\usepackage[]{hyperref}

\newcommand{\ihMpc}{ h\,{\rm Mpc}^{-1}}
\newcommand{\class}{\textsc{\tt CLASS}\xspace}
\newcommand{\camb}{\textsc{\tt CAMB}\xspace}

\begin{document}
\pagestyle{fancy}

\title{Accelerating Large-Scale-Structure data analyses by emulating Boltzmann solvers and Lagrangian \\ Perturbation Theory}
%\titlenote{Emulators to accelerate Large-Scale Structure analyses.}

\author[1,2,3]{Giovanni Aric\`o}
\author[1,4]{Raul E. Angulo}
\author[1]{Matteo Zennaro}
\affil[1]{Donostia International Physics Center (DIPC), Paseo Manuel de Lardizabal, 4, 20018, Donostia-San Sebasti\'an, Guipuzkoa, Spain.}
\affil[2]{Departamento de F\'isica, Universidad de Zaragoza, Pedro Cerbuna 12, 50009 Zaragoza, Spain.}
\affil[3]{Institute for Computational Science, University of Zurich, Winterthurerstrasse 190, 8057 Zurich, Switzerland}
\affil[4]{IKERBASQUE, Basque Foundation for Science, 48013, Bilbao, Spain.}

\maketitle
\thispagestyle{fancy}

\begin{abstract}
The linear matter power spectrum is an essential ingredient in all theoretical models for interpreting large-scale-structure observables. Although Boltzmann codes such as \class or \camb are very efficient at computing the linear spectrum, the analysis of data usually requires $10^4$--$10^6$ evaluations, which means this task can be the most computationally expensive aspect of data analysis. Here, we address this problem by building a neural network emulator that provides the linear theory ({\it total} and {\it cold}) matter power spectrum in about one millisecond with $\approx 0.2\%$ ($0.5\%$) accuracy over redshifts $z \le 3$ ($z \le 9$), and scales $10^{-4} \le k [\ihMpc] < 50$. We train this emulator with more than 200,000 measurements, spanning a broad cosmological parameter space that includes massive neutrinos and dynamical dark energy. We show that the parameter range and accuracy of our emulator is enough to get unbiased cosmological constraints in the analysis of a {\it Euclid}-like weak lensing survey. Complementing this emulator, we train 15 other emulators for the cross-spectra of various linear fields in Eulerian space, as predicted by second-order Lagrangian Perturbation theory, which can be used to accelerate perturbative bias descriptions of galaxy clustering. Our emulators are specially designed to be used in combination with emulators for the nonlinear matter power spectrum and for baryonic effects, all of which are publicly available at \url{http://www.dipc.org/bacco}.

\end{abstract}

% \section{\color{OREblue}Keywords}

% large-scale structure of Universe, cosmological parameters, cosmology: theory, emulator, Boltzmann equations, Lagrangian Perturbation Theory.

\clearpage
\pagestyle{fancy}

\section{Introduction}\label{sec:intro}

The optimal exploitation of large scale structure (LSS) data is one of the most important challenges in modern cosmology. For this, multiple theoretical models have been developed based on perturbation theory, the halo model, excursion set theory, or $N$-body simulations. Regardless of the nature of the modelling, essentially all approaches rely on predictions of the linear matter power spectrum as a function of cosmological parameters.

The linear matter power spectrum can be accurately computed by taking moments of the Bolztmann equation describing the co-evolution of all types of energy (neutrinos, cold dark matter, radiation, \textit{etc.}) in the universe. This truncated Bolztmann hierarchy can be solved very efficiently by publicly available codes such as \textsc{\tt CMBFast} \citep{Seljak:1996}, \camb \citep{Lewis:2000}, and \class \citep{Lesgourgues2011,Blas2011}, which can provide the matter power spectrum in 1--10 seconds of computing, depending on the desired accuracy.

Estimating cosmological parameters from LSS observations typically requires $10^4$--$10^6$ evaluations of the relevant theoretical model. Thus, the computational cost of estimating the linear power spectrum can reach thousands of CPU hours. Traditionally, this has not been an issue because other aspects of the modelling (\textit{e.g}. the calculation of loops in perturbation theory) were significantly more expensive. This situation, however, in changing with the use of {\it emulators}.

Emulators are a mathematical tool that allow for an efficient multidimensional interpolation of a given function. Popular choices in the field of LSS are Gaussian processes \citep[\textit{e.g}.][]{Heitmann2014,McClintock2019b,Bocquet2020}, polynomial chaos expansions \citep[\textit{e.g}.][]{EuclidEmulator,EuclidEmulator2} and neural networks \citep[\textit{e.g}.][]{Kobayashi:2020,Arico2020c,Zennaro2021}. In all cases, a  number of (computationally expensive) predictions at different locations of a parameter space are used to build the emulator which can then be evaluated in any other point of the parameter space.

Multiple emulators have been built for LSS summary statistics (usually based on the results of $N$-body simulations), which include the nonlinear matter power spectrum, halo mass function, halo bias, and the effect of baryons \citep[\textit{e.g}.][]{Heitmann2014,EuclidEmulator,Winther2019,McClintock2019,Zhai2019,Bird2019,Bocquet2020,Schneider2020,Angulo2020,EuclidEmulator2,Arico2020c,Zennaro2021}.

Regardless the technique, the emulated quantity is usually the ratio with respect to the linear theory expectation (\textit{e.g}. the nonlinear over the linear power spectrum). This reduces the dynamical range of the function and removes some of the cosmology-dependence, allowing a more robust and accurate emulation. Since the evaluation of an emulator takes a negligible amount of CPU time, the calculation of the linear power spectrum becomes the bottleneck of the analysis.
Similarly to the emulators, also the widely used \textsc{\tt halofit} algorithm \citep{Smith2003,Takahashi2012} applies its interpolation functions to the linear power spectrum. In this case too, the evaluation of the linear power spectrum is by far the bottleneck of the calculation.

A possible way to tackle this is to reduce the number of evaluations needed in parameter constraints. Specifically, one technique proposed by several authors is to directly emulate the likelihood function \citep[\textit{e.g}.][]{McClintock2019b,Leclercq2018,Pellejero2020}. Another ideas, which we explore here, is to directly speed up the evaluation of the linear power spectrum. %, or the use of neural networks \citep{Manri}

%\teo{I don't follow: one way to tacle this is reduce the number of evaluations. I'd follow with: Another possibility is to make these evaluations extremely fast.}

Even for efficient parameter samplers, reducing the computational cost of Boltzmann solvers at fixed precision would be of great benefit. This would allow a more accurate determination of the respective likelihood functions and a faster exploration of different models and data. An option for speeding up these calculations was proposed by \cite{Albers:2019} who trained a neural network to replace the most time-consuming parts of the code \class. Another alternative is to directly build an emulator for the linear power spectrum. This has been attempted by \textsc{\tt PICO} \citep{Fendt2007} using a fifth-order polynomial expansion to interpolate between pre-computed CMB temperature power spectra, and by \textsc{\tt CosmoNET} \citep{Auld:2007,Auld:2008} using neural networks to emulate CMB fluctuations and the matter power spectrum. Having been developed over a decade ago, these works were limited by the size of their training sets (which contained of the order of 1,000 measurements) and by the computational cost of the training itself.

Here we take advantage of all the recent developments in neural networks and computer architecture to develop a new emulator for the linear matter power spectrum with a focus on the analysis of forthcoming LSS experiments. Our emulator covers a eight-dimensional $\Lambda$CDM parameter space including massive neutrinos and dynamical dark energy, and reaches about 0.2\% precision for $10^{-4} \le k [\ihMpc] < 50$ in a region around the parameter values preferred by current data analyses. The accuracy somewhat degrades to 0.5\% when going to extreme cosmological models. Additionally, we build an emulator for the 15 different cross-spectra that enter a second-order Lagrangian bias expansion of galaxy clustering, which we compute using second-order Lagrangian perturbation theory.

Our neural networks provide predictions in about one millisecond on a single CPU core and have negligible memory requirements. Furthermore, they are designed to be employed together with our emulators for the nonlinear matter power spectrum \citep{Angulo2020}, baryonic effects \citep{Arico2020c}, and galaxy bias expansion \citep{Zennaro2021}. All these emulators are part of the {\tt baccoemu} project, and they are publicly available\footnote{\url{http://www.dipc.org/bacco}} and updated continuously to improve their accuracy.

The outline of this paper is the following: in \autoref{sec:class_emu} we present our linear matter power spectrum and quantify its accuracy; in \autoref{sec:lpt} we introduce our Lagrangian perturbation theory spectra emulators; in \autoref{sec:fit_shear} we employ the linear matter power spectrum emulator to show that it provides unbiased results for the analysis of a mock {\it stage-IV} cosmic shear survey; we provide our conclusions in \autoref{sec:conclusions}.

\begin{table*}
    \centering
    \begin{adjustbox}{max width=1.\textwidth,center}
    \begin{tabular}{ccccccccccccccc} % eight columns, alignment for each
       \hline
       Cosmology &  $A_{\rm s}$ & $n_{\rm s}$ & $\Omega_{\rm c}$ & $\Omega_{\rm b}$ & $h$ & $M_{\rm \nu}\,[{\rm eV}]$ & $w_{0}$ & $w_{\rm a}$ & $z$ \\
       \hline
      Standard  & [-, -] & [-, -] & [0.23, 0.4] & [0.04, 0.06] & [0.6, 0.8] & [0.0, 0.4] & [-1.15, -0.85] &  [-0.3, 0.3] & [0, 3] \\
      Extended  & [-, -] & [-, -] & [0.06, 0.7] & [0.03, 0.07] & [0.5, 0.9] & [0.0, 1.] & [-2., -0.5] &  [-0.5, 0.5] & [0, 9] \\
       \hline
       \end{tabular}
      \end{adjustbox}
      \caption{
The range of cosmological parameter values defining the two hyperspaces over which we train our neural network emulator. $A_{\rm s}$ and $n_{\rm s}$ are the primordial spectral amplitude and tilt, respectively; $\Omega_{\rm c} $ and $\Omega_{\rm b}$ are the density of cold matter and baryons in units of the critical density of the universe; $h$ is the dimensionless Hubble parameter $h= H_0 / (100 \,{\rm km}\,{\rm s^{-1}}{\rm Mpc^{-1}})$; $M_{\rm \nu}$ is the sum of neutrinos masses in eV; and $w_0$ and $w_{\rm a}$ are parameters describing the time-evolving dark energy equation of state via $w(z) = w_0 + (1-a)\,w_{\rm a}$; $z$ is the redshift.
      }
    \label{tab:cosmologies}
\end{table*}

\section{Linear matter power spectrum emulator}\label{sec:class_emu}

In this section, we describe our emulator for the linear cold matter power spectrum. We start by defining our parameter space (\autoref{sub:par_space}), then we describe our training and validation sets (\autoref{sub:train_val}), as well as our neural network setup (\autoref{sub:nn}). We finish by validating our predictions for the growth function, baryonic acoustic oscillations, and the neutrino-induced suppression on the power spectrum (\autoref{sub:validation_emu}).

\subsection{Parameter space}\label{sub:par_space}

We consider 8 cosmological parameters: the primordial spectral amplitude $A_{\rm s}$ and index $n_{\rm s}$; the density of cold matter and baryons in units of the critical density of the universe, $\Omega_{\rm c} $ and $\Omega_{\rm b}$, respectively; the dimensionless Hubble parameter, $h \equiv H_0 / (100 \,{\rm km}\,{\rm s^{-1}}{\rm Mpc^{-1}})$; the sum of neutrinos masses in units of eV, $M_{\rm \nu}$; and a time-evolving dark energy equation of state defined through a CPL parameterisation \citep{ChevallierPolarski2001,Linder2003}, $w(z) = w_0 + (1-a)\,w_{\rm a}$; the redshift $z$.

We define two separate cosmological hyperspaces. The first one, hereafter dubbed as {\it standard}, spans values roughly $10\sigma$ around Planck best-fitting parameters \citep{Planck2018}. We note that this hyper-volume is very similar to that adopted in the nonlinear, baryonic, and bias emulators of \cite{Angulo2020,Arico2020c, Zennaro2021}, respectively. The second cosmological parameter space, dubbed as {\it extended}, is defined by parameter ranges roughly twice as large, with which we aim at expanding the possible usage of the emulator to different kind of analyses and applications. The cosmological parameters and the respective ranges of both hyperspaces are provided in \autoref{tab:cosmologies}. We note that, since the dependence of the power spectrum on $A_{\rm s}$ and $n_{\rm s}$ is straightforward at linear level, we can avoid to train the neural network for these two parameters. Thus, we train the network with fixed $A_{\rm s}'$ and $n_{\rm s}'$, and then rescale the power spectrum as follow:
\begin{equation}
P(k, A_{\rm s}, n_{\rm s}) = \frac{A_{\rm s}'}{A_{\rm s}} \left( \frac{k}{k_{\rm p} h^{-1}} \right)^{n_{\rm s}-n_{\rm s}'} P(k, A_{\rm s}', n_{\rm s}'),
\end{equation}
whit a pivot scale $k_{\rm p}=0.05 \, {\rm Mpc}$. In this way, the neural network is facilitated by a lower dimensionality problem, we can in principle use less points in the training set, and additionally we are not limited by boundaries in $A_{\rm s}$ and $n_{\rm s}$\footnote{we note that this procedure is practically equivalent to the emulation of the power spectrum transfer function.}.

\subsection{Training and validation sets}\label{sub:train_val}

We define our training set by evenly sampling the parameter space with a Latin-hypercube (LH) algorithm, a statistical method which maximise the distance between the sampling points. Specifically, our neural networks are trained with the co-addition of various LH samples. The first one is a 50,000-point LH defined in the {\it standard} cosmological space. The second one is a LH of 100,000 points, defined in the {\it extended} cosmological space. We then added two LHs, of 10,000 and 20,000 points, defined respectively in the {\it standard} and {\it extended} space, but fixing the redshift at $z=0$. Two extra LHs of 10,000 and 20,000 points are defined in the {\it standard} and {\it extended} space but within a $\Lambda$CDM cosmology, \textit{i.e}. fixing $M_{\rm \nu}=0.$, $w_0=-1$, and $w_{\rm a}=0$. Finally, the last two LHs, also of 10,000 and 20,000 points, are built in the {\it standard} and {\it extended} space, within a $\Lambda$CDM cosmology and at $z=0$. In this way, we improve our predictions for the $\Lambda$CDM case and at $z=0$, both of which of particular cosmological interest but that might display degraded performances since they are located on the edge of the parameter hyperspace.

We note that the {\it standard} space is more densely sampled, and thus we expect our emulator to be more accurate in this region.  We use 90\% of the points in each LH in the training set, and add the remaining 10\% to the validation set.

Overall, our training set is made of 216,000 models, whereas our validation set contains 24,000. We employ the Boltzmann solver \class to compute the cold-matter power spectrum in each of the 240,000 points. We make our predictions on a fixed grid in wavenumber, with $600$ $k-$bins between $10^{-4} \le k [\ihMpc]  \le 50$, and at the redshift indicated by the LH sampling. In \autoref{app:setup} we provide details of our specific \class setup and a comparison with \camb.
We highlight that we emulate both the power spectrum of cold dark matter plus baryons (\textit{i.e}. {\it cold} matter), and the total mass case (including massive neutrinos when present). We made this choice since the cold matter power spectrum has been shown to be a better prediction for the various LSS statistics \citep[\textit{e.g}.][]{Castorina:2015,Zennaro:2019}, but the total matter power spectrum can be useful, for example, for weak lensing analyses.

Before the training, we take the logarithm of the power spectra, normalising them by their mean in each $k$-bin. We find that, for the size of our training set, this is sufficient to reduce the variance of the data, and that using the ratio of the power spectra with some approximated method, \textit{e.g}. \cite{EisensteinHu1999}, does not improve sensibly the emulation. We have also tried to split the training set in two different components and emulating them separately, \textit{e.g}. power spectrum at $z=0$ and growth function, or smooth power spectrum and BAO oscillations, but we did not find any obvious advantage in any of these strategies.

Finally, we perform a principal component analysis (PCA) decomposition and retain the first 64 eigenvectors, which combined are enough to reproduce the broad band power spectrum below 0.03\% and the BAO below 0.1\%; this effectively filters out small scale noise, which aids the neural network training. We note that, in order to get the BAO accuracy below 0.1\%, more PCs have to be included, at the price of a more complex training of the network.

%============================================================

\begin{figure*}
\begin{center}
\includegraphics[width=.6\textwidth]{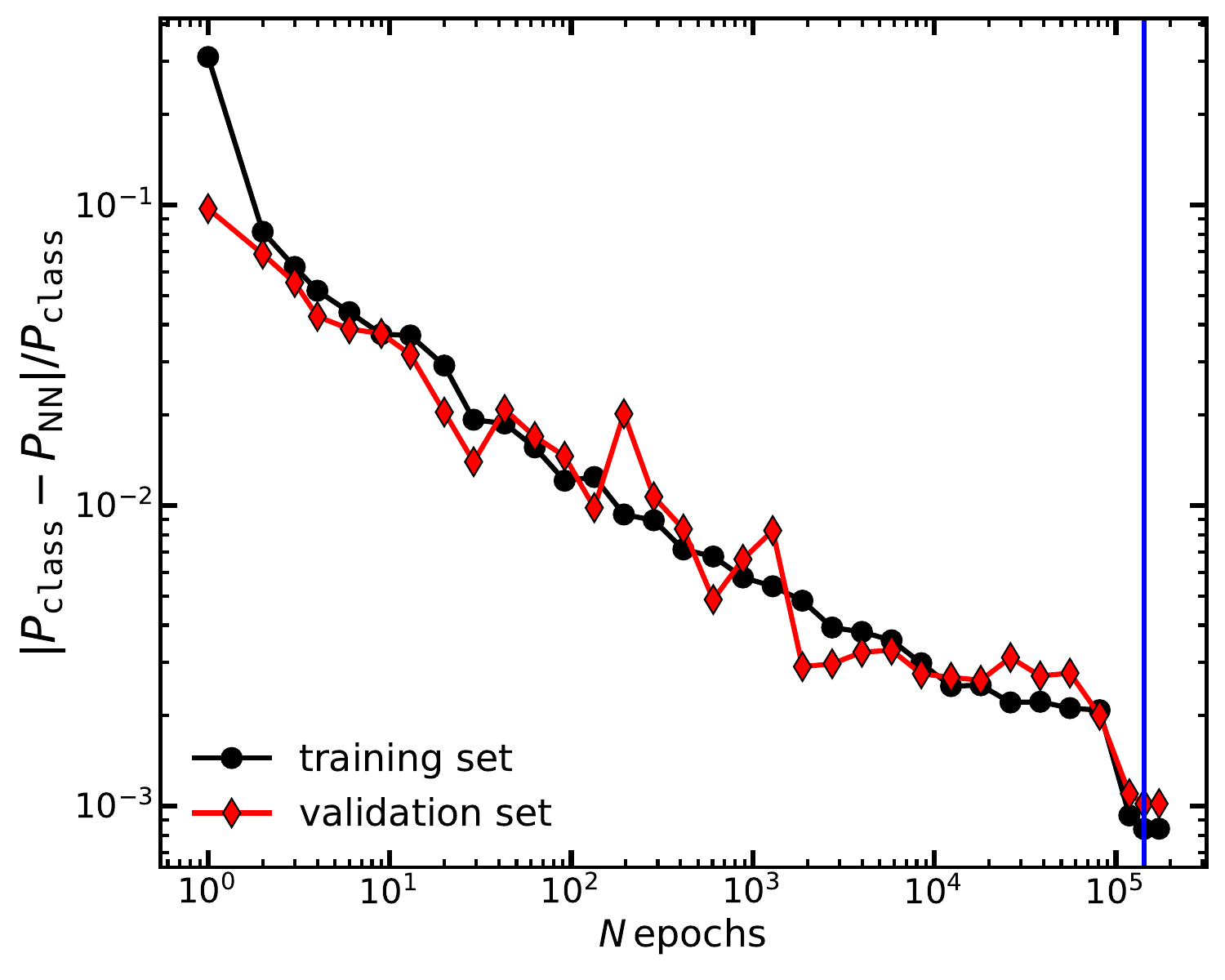}
\caption{The mean absolute fractional error of our neural network, $\lambda = \left \langle \frac{|P_{\rm NN}-P_{\class}|}{P_{\class}} \right \rangle$, as a function of the number of epochs employed for its training. Black circles and red diamonds show the results when $\lambda$ is evaluated on the training and the validation set, respectively. The vertical blue line marks the minimum of $\lambda$ in the validation set and thus the training we will adopt thereafter in this paper. }
\label{fig:training}
\end{center}
\end{figure*}
%============================================================

%============================================================
\begin{figure*}
\begin{center}
\includegraphics[width=1.\textwidth]{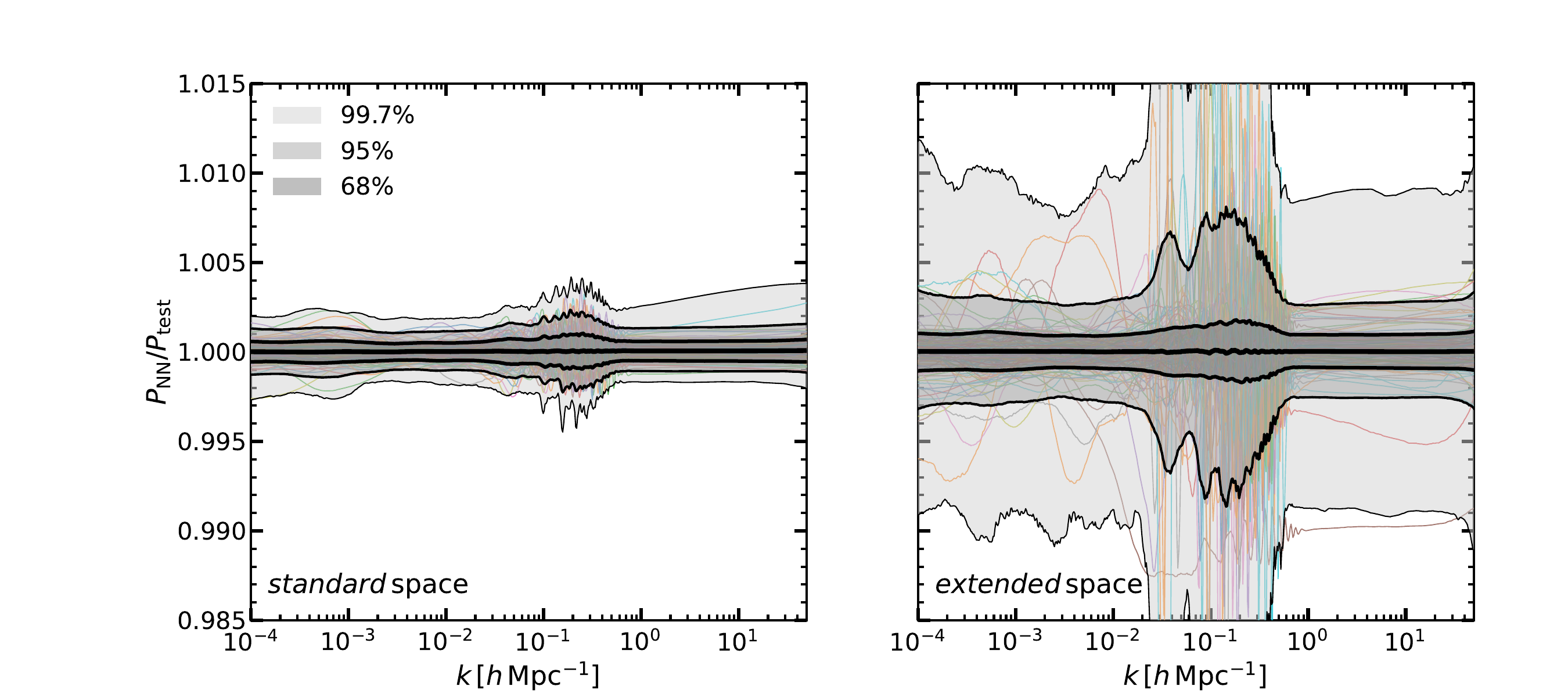}
\caption{The accuracy of our neural network predictions for the linear matter power spectrum for multiple cosmologies and redshifts. We display the ratio of the power spectra computed by our emulator, $P_{\rm NN}$, to that of the Boltzmann solver \class, $P_{\class}$, in the {\it standard} (left panel) and in the {\it extended} (right panel) cosmological parameter spaces (see \autoref{tab:cosmologies}).
In the {\it extended} space, {\it all} the cosmological parameters have values not included in the {\it standard} space.
The shaded regions enclose 68\%, 95\%, and 99.7\% of the cosmologies in our validation set, and the mean is shown as a thick black line. As an example, thin coloured lines show the results for 100 randomly selected cosmologies.}
\label{fig:accuracy}
\end{center}
\end{figure*}
%============================================================

\subsection{Neural network setup}\label{sub:nn}

We build our emulator using a feed-forward neural network trained with 216,000 power spectrum measurements. Compared to Gaussian Processes, neural networks have the advantage of significantly better scaling of the computational and memory requirements with large training sets. We employ the publicly-available libraries \textsc{\tt Keras} and \textsc{\tt TensorFlow} \citep{chollet2015keras,abadi2016tensorflow} to build a relatively simple architecture with two hidden layers of $400$ neurons each, and a rectified linear unit as an activation function. We use an Adam optimiser with an initial learning rate of $10^{-3}$, and define as loss function the mean absolute fractional error. This quantity, to be minimised during the training procedure, is defined as

\begin{equation}
\lambda = \left\langle \frac{|P_{\rm NN}-P_{\class}|}{P_{\class}} \right\rangle,
\end{equation}

\noindent where $P_{\rm NN}$ is the power spectrum predicted by the neural network, $P_{\class}$ is the \class power spectrum, and the mean $\langle ... \rangle$ runs over the training points.

To avoid overfitting, we monitor the loss function $\lambda$ computed over both the training and the validation datasets. We stop the training when the loss function evaluated on the validation set does not decrease for more than 10,000 epochs (to avoid being stuck in local minima). We then reduce the learning rate by a factor of 10 and repeat the training until the loss function becomes flat again. This resulted into a training of approximately $10^5$ epochs, being the final loss function of the order of $10^{-3}$. The whole training took approximately 24 hours. We apply this procedure twice, one for the total matter power spectrum and one for the cold matter one.

In \autoref{fig:training} we display the loss function $\lambda$ computed in the training and validation sets, as a function of the number of epochs employed in the training of the total matter power spectrum. We see that the neural network return progressively more accurate results, with the loss function decreasing roughly as $N_{\rm epoch}^{-0.2}$. We can also see that the accuracy is roughly identical in the training and validation sets, which suggests the network has learnt relevant features in the power spectrum data, rather than any specific source of noise. This trend is qualitatively similar for the cold matter power spectrum.
We note that it appears that further training could yield to an even more accurate neural network without overfitting. However, as the improvement scales slowly with $N_{\rm epochs}$, it becomes impractical to train for much longer. Furthermore, the accuracy of the network is comparable with the residuals of the PCA decomposition and with the level of agreement between \class and \camb that we observe in \autoref{app:setup}.

%============================================================
\begin{figure*}
\begin{center}
\includegraphics[width=.33\textwidth]{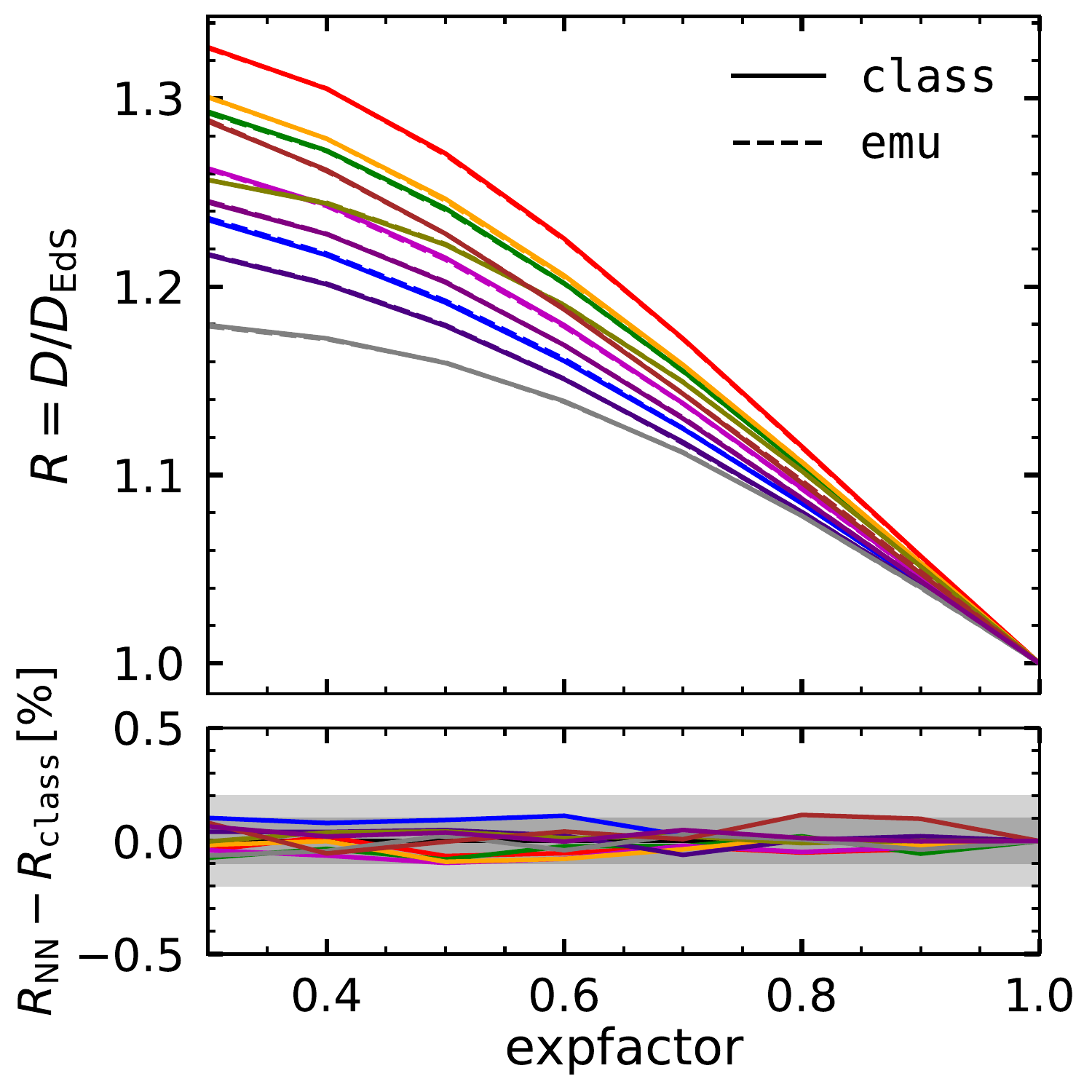}
\includegraphics[width=.33\textwidth]{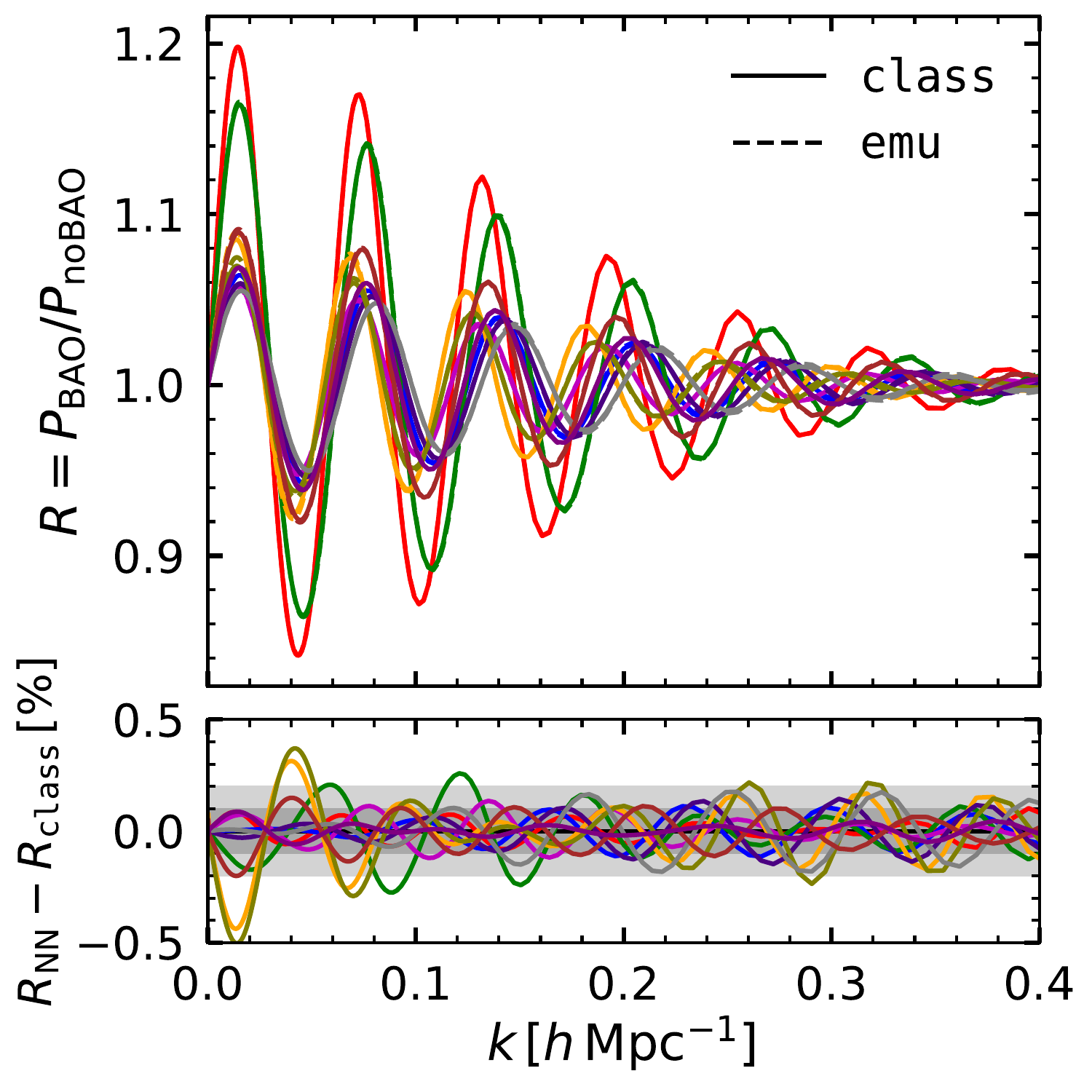}
\includegraphics[width=.33\textwidth]{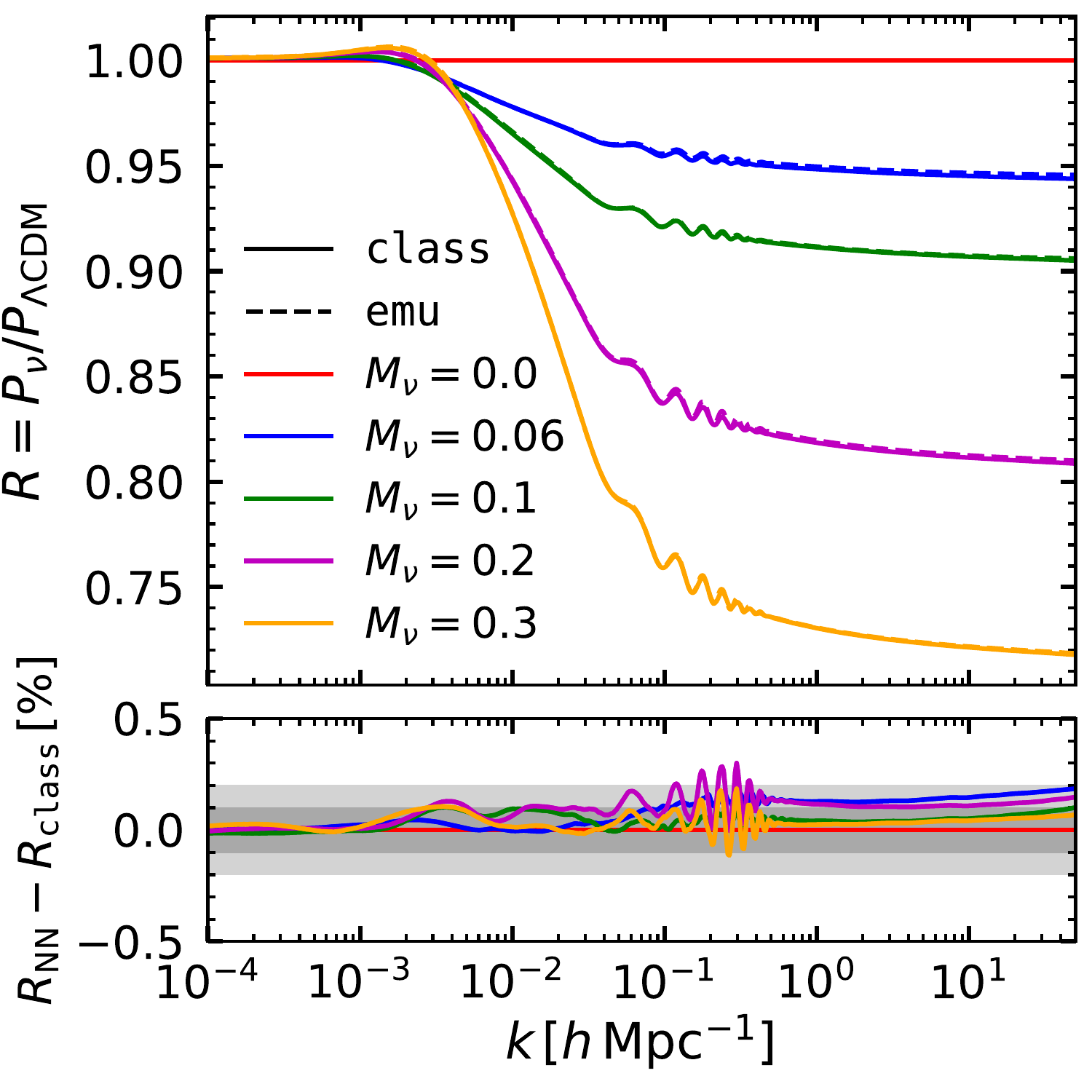}
\caption{Validation of the predictions of our neural-network emulator for the linear matter power spectrum. {\it Left panel:} Growth factor at $k=0.2\,\ihMpc$ as a function of expansion factor. {\it Middle panel:} The ratio of the linear power spectrum over its smooth (or de-wiggled) counterpart which isolates the contribution of baryonic acoustic oscillations to the power spectrum. In both of these panels we show 10 randomly-selected cosmological models within the {\it standard} space. {\it Right panel:} The ratio of the power spectrum computed including  neutrinos of a various mass, as specified in the legend, over its respective neutrino massless case. In all three cases, we show the results obtained with \class as solid lines, and with our emulator as dashed lines. In the bottom panels, we display the compare these predictions indicating differences of 0.1\% and 0.2\% as shaded regions.}
\label{fig:accuracy_three}
\end{center}
\end{figure*}
%============================================================

\subsection{Validation of the emulator}\label{sub:validation_emu}

We test the accuracy of our emulator using the validation set described in \autoref{sub:train_val}. We recall that this sample contains roughly $10\%$ of the training data, distributed between the {\it standard} and {\it extended} cosmological spaces.

In \autoref{fig:accuracy} we display the ratio of the total matter power spectrum computed with our emulator to that of \class. Shaded regions contain 68, 95, and 99.7\% of the measurements in the {\it standard} and {\it extended} space (left and right panel, respectively). We display 100 randomly-selected ratios for comparison.

We can see that, in the {\it standard} cosmological space, our emulator is unbiased at the $0.01\%$ level, and 68\% of the validation set lies within 0.1\%, while 95\% within 0.2\%. Outliers appears to be well within 0.4\%. In the {\it extended} cosmological space, the accuracy is about $\approx 0.3\%$ for most of the scales, although with more outliers, especially around the BAO scales. Nevetheless, we highlight the {\it standard} cosmological space is expected to contain the full range of currently allowed cosmologies. In any case, even in the {\it extended} space, our emulator accuracy is significantly higher than that of current predictions for the nonlinear matter power spectrum for which state-of-the-art $N$-body codes agree at $\sim2\%$ for $k\sim10\ihMpc$ \citep{Schneider2016,Angulo2020,Springel:2020}. Moreover, in general we expect in the {\it extended space} an accuracy in-between the one found in the {\it standard} and {\it extended} space, being the latter a very particular case when {\it all} the parameters simultaneously have very extreme values.
Therefore, any uncertainty of our emulator is arguably subdominant for LSS data analyses.

We further test our total matter power spectrum emulator in \autoref{fig:accuracy_three} by examining its predictions for the redshift dependence of growth factor, the baryonic acoustic oscillations, and the neutrino-induced suppression of the matter power spectrum. In each case we illustrate the accuracy by displaying the emulator and \class predictions for a small number of selected cosmologies within our {\it standard} space.

In the left panel of \autoref{fig:accuracy_three} we show the linear growth factor, $D(k) \equiv \sqrt{P(k,z)/P(k,z=0)}$ over the Einstein-de-Sitter solution, $D \propto a$, computed using our emulator and \class at 10 different cosmological models. We see that the emulator predictions are very accurate, within 0.1\%, as expected from the performance shown previously. Also, the accuracy does not depend on the expansion factor, which indicates that the performance of our emulator is the same at all redshifts.

Given their importance for LSS analyses, we explicitly check how well the baryonic acoustic oscillation (BAO) feature is recovered. In the middle panel of \autoref{fig:accuracy_three} we display the ratio between linear power spectrum and its de-wiggled counterpart, $P/P_{\rm no-wiggle}$, which highlights the contribution of the BAO to the power spectrum. We compute $P_{\rm no-wiggle}$ by performing a discrete sine transform of the linear theory power spectrum, smoothing the result, and returning to Fourier space with an inverse sine transform \citep[\textit{e.g}][]{Baumann:2018}. We can see how each oscillation is remarkably well reproduced by our emulator, with small in-phase deviations of the order of $0.3\%$. This suggests the accuracy of our emulator is also high enough for BAO analyses.

To close this section, we examine the dependency of the power spectrum on massive neutrinos. The rightmost panel of \autoref{fig:accuracy_three} shows the ratio of the power spectrum computed with increasingly massive neutrinos, $M_{\rm \nu} = [0.06,0.1,0.2,0.3]$ eV, over the massless case, $M_{\rm \nu} = 0$ eV, keeping fixed the primordial spectral amplitude $A_{\rm s}$ and the matter density $\Omega_{\rm m}$. As with our previous tests, we find that the  distortion caused by neutrinos, a suppression at small scales proportional to the neutrino mass fraction, is accurately recovered by our emulator at $0.2\%$ level.

In this section, we have shown the tests carried out to validate the accuracy of our total matter power spectrum emulator. We have performed similar tests for the cold matter power spectrum emulator, which we do not include here for the sake of brevity, finding a similar level of accuracy.

\begin{figure*}
\begin{center}
\includegraphics[width=.99\textwidth]{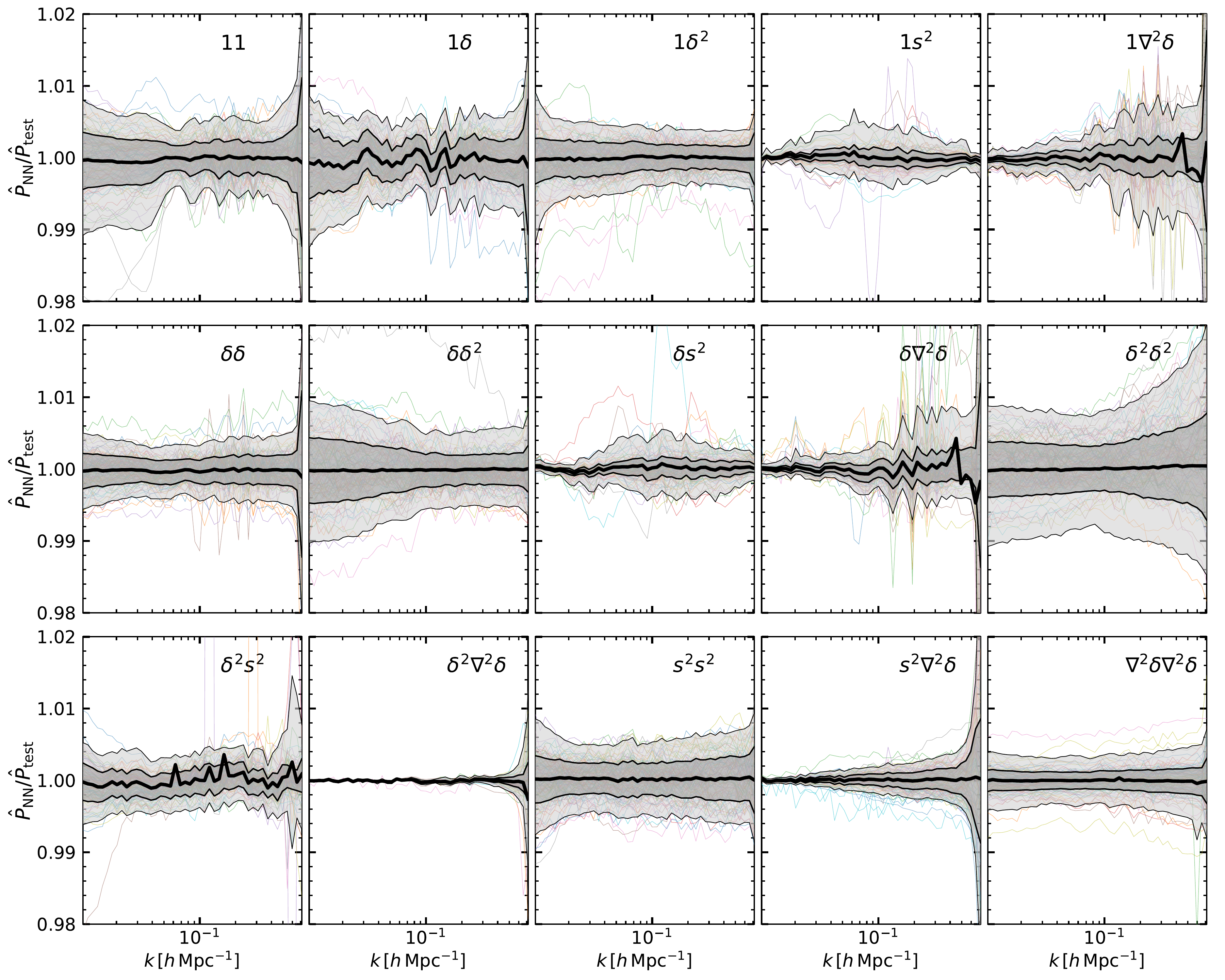}
\caption{The accuracy of our emulators for the cross-spectrum of linear fields in Eulerian coordinates predicted by Lagrangian Perturbation Theory. The two fields defining the cross-spectra are indicated in the legend of each panel, where $1$ is an homogeneous Lagrangian field; $\delta$ and $\delta^2$ are the linear density field and its square, respectively; $s^2$ is the shear field; and $\nabla^2\delta$ is the Laplacian of the linear density field. In each panel we display the ratio of the emulator prediction over the same quantity computed by directly solving the relevant LPT expression. Shaded regions enclose 68\% and 95\% of the measurements in our validation set, and the mean is is marked by the thick black line. For comparison we show a randomly-selected set of cosmologies as coloured lines.}
\label{fig:models_lpt}
\end{center}
\end{figure*}

\section{Lagrangian perturbation theory emulator}\label{sec:lpt}

Besides predicting the clustering of matter, another challenge for present-day and forthcoming cosmological surveys is to predict the clustering of galaxies. In this section we build emulators for several matter statistics as predicted by Lagrangian perturbation theory (LPT) that are typically employed in models for the power spectrum of galaxies.

\subsection{Lagrangian bias expansion}

Among the many possible models to describe the clustering of galaxies, a particularly promising one is a perturbative Lagrangian bias expansion \citep[][]{Matsubara2008}. In this formalism, the power spectrum of galaxies, $P_{\rm gg}$, and the matter-galaxy cross-power spectrum, $P_{\rm gm}$, at second order are given as:

\begin{equation}
    \begin{split}
        P_{\rm gg} &= \sum_{i,j \in \{1,\delta,\delta^2,s^2,\nabla^2\delta\}} b_i b_j\,P_{ij},\\
        P_{\rm gm} &= \sum_{i \in \{1,\delta,\delta^2,s^2,\nabla^2\delta\}} b_i \,  P_{i1},  \\
\end{split}
\end{equation}

\noindent where, $b_{\delta}, b_{\delta^2}, b_{s^2},$ and $b_{\nabla^2\delta}$ are free  ``bias'' parameters; $\delta$ is the linear density field, $s^2$ is the shear field defined as $s^2 \equiv s_{ij}s_{ij}$ where $s_{ij} \equiv \partial_i \partial_j \nabla^{-2}\delta - \delta_{D,ij} \delta(q)$, and $\nabla^2 \delta$ the Laplacian of the linear density; and $P_{ij} = \langle |\delta_i(\boldsymbol{k}) \delta_j^*(\boldsymbol{k})|\rangle$    \cite[see][for details]{Zennaro2021}.

In other words, the clustering of galaxies is given as a weighted sum of 15 cross-spectra of five Lagrangian fields, $\delta^{\rm L} \in \{1,\delta,\delta^2,s^2,\nabla^2\delta\}$, advected to Eulerian coordinates:

\begin{equation}
1+\delta^{\rm E}(\boldsymbol{x}) = \int \mathrm{d}^3 q \, \delta^{\rm L}(\boldsymbol{q}) \delta_{\rm D}(\boldsymbol{x}-\boldsymbol{q}-\boldsymbol{\Psi}(\boldsymbol{q}))
\end{equation}

\noindent where $\delta_{\rm D}$ is a Dirac's delta. The displacement field $\Psi(\boldsymbol{q})$ (and thus the cross-spectra) can be computed perturbatively at a given order, or, as proposed recently, measured directly from $N$-body simulations. The latter can accurately describe the power spectrum of galaxies down to much smaller scales than using perturbative solutions, which opens up the possibility of an accurate modelling of galaxy clustering \citep{Modi2020}.

In fact, \cite{Zennaro2021} and \cite{Kokron2021} have built emulators for these spectra as a function of cosmology from simulation suites. However, even in this case, large scales are  described with perturbation theory owing to the noise (cosmic variance) in the $N$-body displacements. Additionally, the emulation is performed for the ratio of the $N$-body spectra over the perturbative solution, which, as for the matter power spectrum, improves the quality of the emulation.

The LPT terms can be computed with publicly available codes \citep[such as {\tt velocileptors},][]{Chen2020} or, as in \cite{Zennaro2021}, by solving the three dimensional LPT integrals with a adaptive quadrature algorithm. Both implementations are very efficient and produce the LPT spectra in around 1 second of computing time, depending on accuracy parameters and machine architecture. However, this can become the most costly step in the context of obtaining the galaxy model from an emulator.

In this section, we build an emulator for each of the 15 LPT terms to speed up this process. Specifically, we emulate the terms $P_{11}, P_{1\delta},$ and $P_{\delta\delta}$ where we expand $\delta$ at first order in Lagrangian perturbation theory; this is enough for the applications mentioned. All remaining terms are computed expanding densities at second order in Lagrangian perturbation theory, retaining only contributions of order (11) and (22). We refer the reader to \cite{Zennaro2021} for the explicit expression of each of these spectra in LPT. It is important to note that we assume the linear power spectrum entering our LPT calculations to be smoothed on a scale $k_{\rm s} = 0.75 h \, \mathrm{Mpc}^{-1}$.

\begin{figure*}
\begin{center}
\includegraphics[width=.93\textwidth]{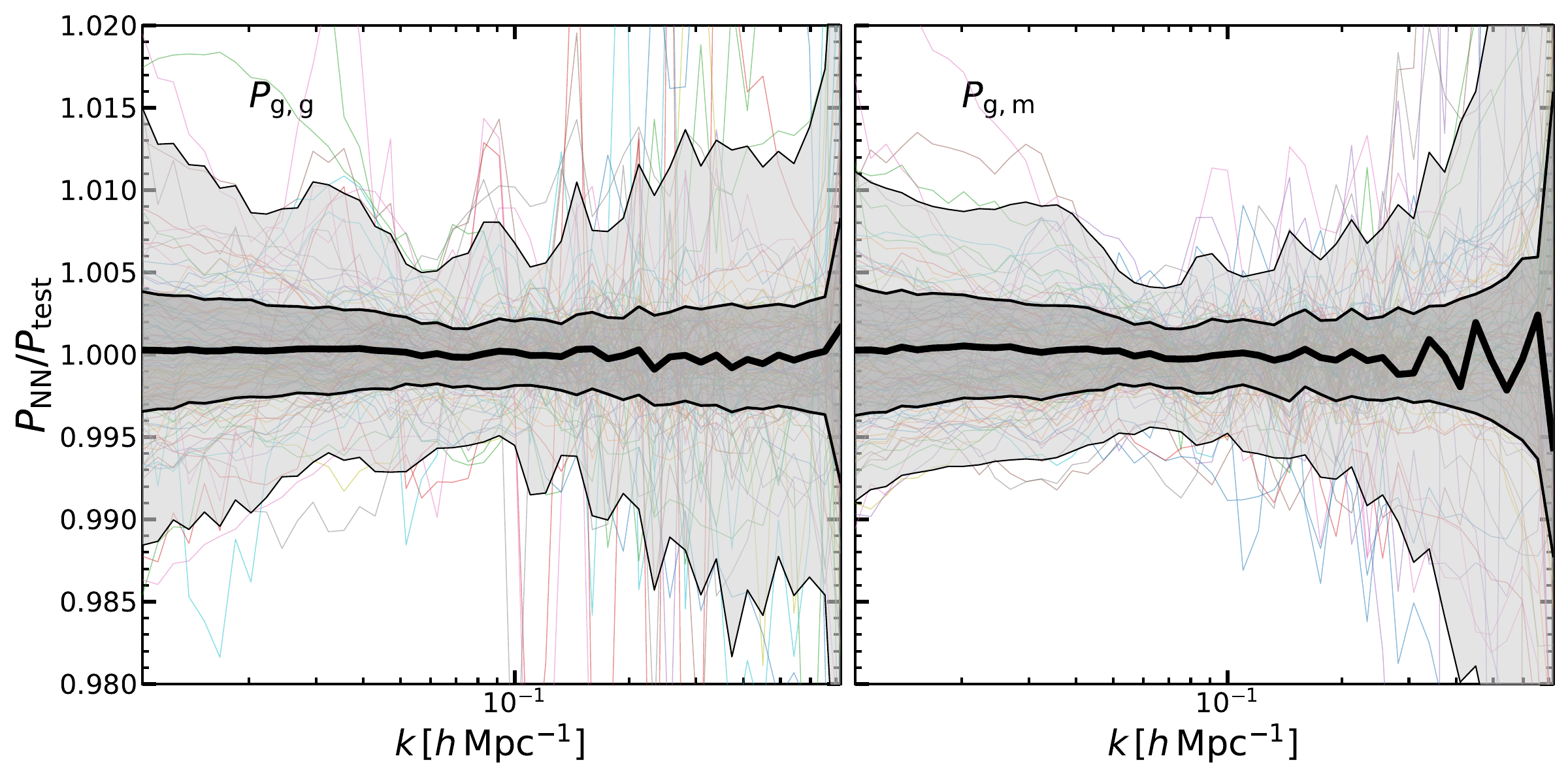}
\caption{The accuracy of our emulators for the galaxy auto power spectrum (left panel) and galaxy - matter cross-spectrum predicted by Lagrangian Perturbation Theory. We used bias factors randomly drawn from the priors computed in \cite{Zennaro2021}. In each panel we display the ratio of the emulator prediction over the same quantity computed by directly solving the relevant LPT expression. Shaded regions enclose 68\% and 95\% of the measurements in our validation set, and the mean is is marked by the thick black line. For comparison we show a randomly-selected set of cosmologies as coloured lines.}
\label{fig:galaxy_lpt}
\end{center}
\end{figure*}

\subsection{LPT emulator}

We build the LPT emulators in the {\it standard} hyper-parameter space (see \autoref{tab:cosmologies}) and over the expansion factor range: $a \in [0.4, 1.]$ (note that this redshift range is slightly smaller than that employed in \autoref{sub:par_space}). We note that our cosmological parameter space matches that used by \cite{Zennaro2021}, where we presented non-linear boost-factors for these spectra. We then compute the LPT predictions in 46 logarithmic bins over the range $10^{-2} < k [\ihMpc] < 0.75$.

We sample the parameter space with a single LH with 20,000 points. In each of these points, we compute the 15 Lagrangian cross power spectra as described above. As noted by \cite{Zennaro2021}, some of these spectra can be negative. In the case where a term has at least a negative value in the training set, we emulate the quantity:

\begin{equation}
\hat{P}_{ij} = P_{ij} + |\min(P_{ij})|+0.1.
\end{equation}

\noindent so that the spectra are always positive. As in the case of the linear matter power spectrum, prior to the training we take the logarithm of the power spectra, and subtract the mean in each $k$-bin.  We then perform a PCA decomposition prior to the training, retaining for each spectrum a number of PC sufficient to recover the power spectrum at $0.01\%$, which ranges from five to 20 PCA vectors.

We use 90\% of the sample as our training dataset and the remaining 10\% as our validation set. The architecture of the neural network used to emulate the LPT power spectra is the same as that described in \autoref{sub:nn}, \textit{i.e}. 2 hidden layers with 400 neurons each.

We quantify the accuracy of our emulators in \autoref{fig:models_lpt}, which shows the ratio of the emulation prediction in the validation set over the spectra computed with LPT. Overall, the accuracy of the 15 spectra is better than $1\%$, with some terms ($1\delta^2$, $1s^2$, $\delta\delta$, $\delta s^2$, $\delta^2\nabla^2\delta$, $s^2\nabla^2\delta$, and $\nabla^2\delta\nabla^2\delta$) more accurate than $0.5\%$. 

When computing the power spectrum of a biased tracer (e.g. galaxies), the 15 emulated cross-spectra are weighted by the respective bias factors, and thus have different contribution to the total spectrum. To quantify the accuracy of the emulation in the case of realistic galaxies, we randomly sample the priors on Lagrangian galaxy bias found in \cite{Zennaro2021}. \autoref{fig:galaxy_lpt} shows the accuracy we get in the galaxy auto power spectrum and in the galaxy - matter cross power spectrum, that is around 1\% at most of the scales considered, and approaching 2\% at scales $k > 0.4 \, \ihMpc$. 

This accuracy is higher than that of the nonlinear emulators, thus the LPT emulation uncertainty should add a subdominant contribution to the global modelling error. Nevertheless, as with our linear emulator, it is straightforward to add further points to the training set, if higher accuracy is required.

\section{Parameter constraints from mock cosmic shear power spectra}\label{sec:fit_shear}

\begin{figure*}
\includegraphics[width=.9\textwidth]{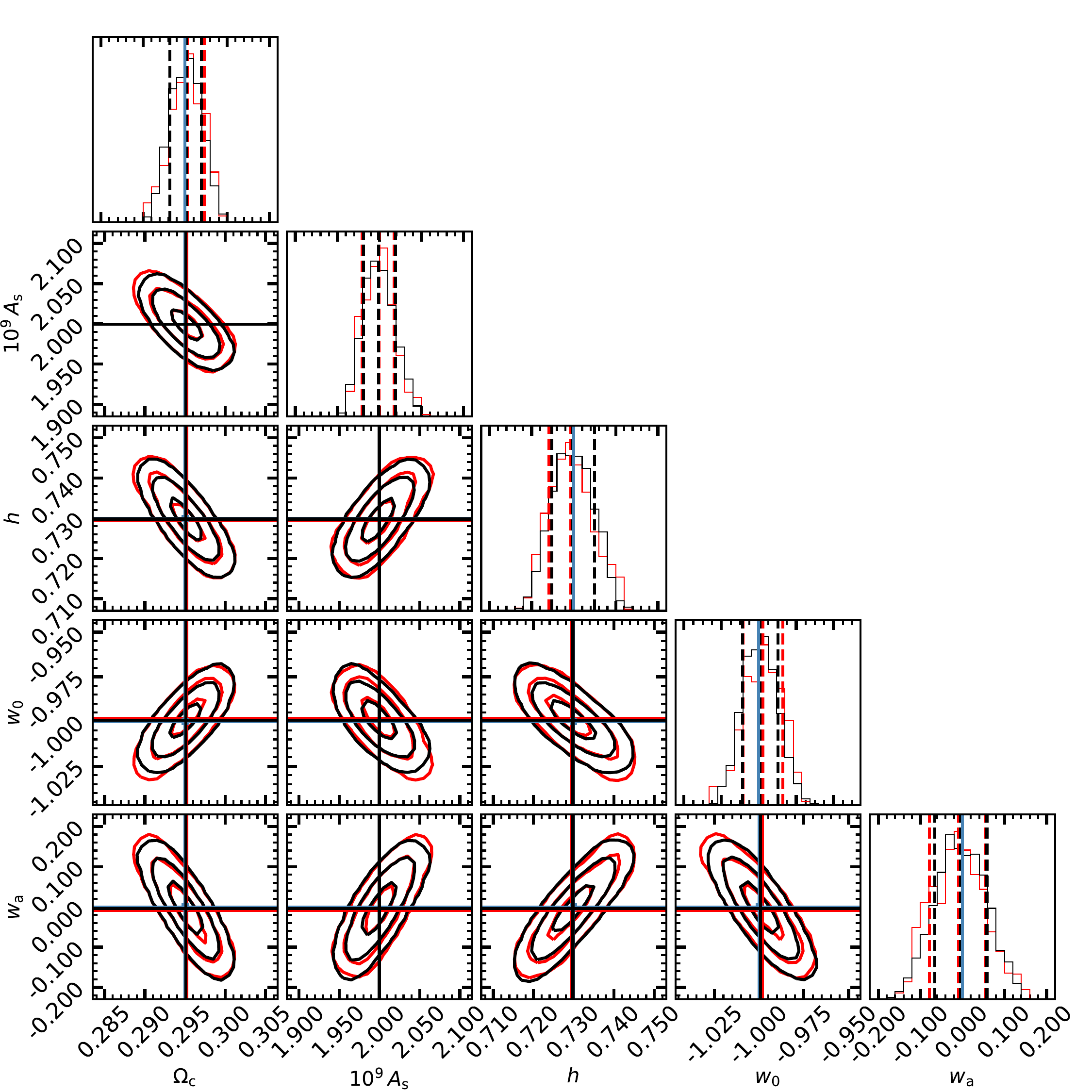}
\caption{The marginalised posterior distributions on cosmological parameters  obtained from mock weak lensing data. Our mock data is comprised by the shear power spectrum and cross-spectra using 10 tomographic bins over $z \in [0.,2.5]$, and a redshift distribution of background galaxies, shape noise, and cosmic variance expected for a {\it stage-IV} Euclid-like survey. The contours in each panel show 1,2, and $3~\sigma$ levels obtained by performing an MCMC analysis using the linear matter power spectrum directly provided by the Boltzmann solver \class (black lines and contours) or by our emulator (red lines and contours). Blue lines indicate the  cosmological parameters adopted in our mock data, whereas black and red lines show the best-fitting values obtained by our analysis.}
\label{fig:contours}
\end{figure*}

In this section, we illustrate the use of our linear matter power spectrum emulator with a simple application: we infer cosmological parameters from a mock lensing power spectrum using either \class or our emulator. In this way, we will confirm that the accuracy of our emulator is adequate in the context of LSS data analysis.

Our mock data corresponds to the auto and cross power spectra of weak lensing shear measurements for 10 equi-populated redshift bins $z_i \in [0.1, 2.5]$. We adopt the specifics of a {\it stage-IV} survey, and in particular those of the Euclid mission \citep{Amendola2018}. The cross-spectrum of cosmic shear is given by:

\begin{equation}
\label{eq:cl}
C_{\gamma_i, \gamma_j}(\ell) = \int_{0}^{\chi_{\rm H}}  \frac{g_i(\chi) g_j(\chi)}{\chi^2} P \left( \frac{\ell}{\chi}, z(\chi) \right) d\chi,
\end{equation}

\noindent where $P(k,z)$ is the linear (total) matter power spectrum (given by \class for our mock data), and $g_i(\chi)$ is the the lensing kernel of the $i$-th redshift bin:

\begin{equation}
g_i(\chi) = \frac{3}{2}\Omega_{\rm m} \left( \frac{H_0}{c} \right)^2  \frac{\chi}{a} \int_{z(\chi)}^{z_{\rm H}} {\rm d}z^\prime \, n_i(z) \frac{\chi(z^\prime)-\chi(z)}{\chi(z^\prime)},
\end{equation}

\noindent where $c$ is the light speed and $\chi(z)$ is the comoving distance to $z$. We assume a redshift distribution of galaxies in the $i$-th bin $n_i(z) \propto z^2 \exp \left( - \frac{z \sqrt{2}}{z_i} \right)$, with $z_{i}$ roughly setting the redshift where the galaxy number density peaks. The galaxy distribution in each bin is normalised such that $\int_{0}^{\infty} {\rm d}z\,n_i(z) = 1$.

For simplicity, we only consider the Gaussian contribution to the covariance:

\begin{multline}
{\rm Cov}^{\rm G}_{\gamma_i,\gamma_j,\gamma_m,\gamma_n} (\ell_1, \ell_2) = \frac{\delta_{\ell_1,\ell_2}}{N_{\ell}} \Bigg[
\left( C_{\gamma_i,\gamma_m} + \delta_{im} \frac{\sigma_{\rm e}^2}{2\bar{n}_{\rm eff}^i} \right)   \left( C_{\gamma_j,\gamma_n} + \delta_{jn} \frac{\sigma_{\rm e}^2}{2\bar{n}_{\rm eff}^j} \right) +
\left( C_{\gamma_i,\gamma_n} + \delta_{in} \frac{\sigma_{\rm e}^2}{2\bar{n}_{\rm eff}^i} \right) \left( C_{\gamma_j,\gamma_m} + \delta_{jm} \frac{\sigma_{\rm e}^2}{2\bar{n}_{\rm eff}^j} \right) \Bigg]
\label{eq:covariance}
\end{multline}

\noindent where $N_{\ell} = (2\ell +1) \Delta \ell f_{\rm sky} $ is the number of independent multipoles falling in a bin centered in $\ell$ with width $\Delta \ell$, $f_{\rm sky}$ is the fraction of the sky covered by the survey, $\sigma_{\rm e}$ and $\bar{n}_{\rm eff}^i$ are the RMS ellipticity and the effective projected number density of the source galaxies in the $i$ bin, respectively \citep{Barreira2018}. The presence of the Kronecker deltas $\delta_{\ell_i,\ell_j}$ forces the Gaussian term to be diagonal, and $\delta_{ij}$ to have shape noise terms only for matching redshift bins.
We adopt a fiducial setup of $f_{\rm sky}=0.36$, $\sigma_{\rm e}=0.37$, $\bar{n}_{\rm eff}^i=30 \, {\rm arcmin}^{-2}$.

We fit the mock data over 20 logarithmically-spaced multipoles between $\ell \in [20,5000]$, using the affine invariant MCMC sampler {\tt emcee} \citep{Foreman2013} and employing 10 walkers of 10,000 steps each, considering a burn-in phase of 1,000 steps. Our data model is given by \autoref{eq:cl} computed with a linear power spectrum provided by either \class or by our emulator, and letting free the cosmological parameters $\Omega_{\rm c}$, $\sigma_8$, $h$, $w_0$, and $w_{\rm a}$. Using 1 CPU, a single call to \class takes $\approx 0.5$ seconds ($\approx 7$ seconds when having massive neutrinos), whereas the emulator evaluation takes $\approx 1$ milliseconds.

In \autoref{fig:contours}, we display the posterior distributions estimated with the MCMC analysis using \class and the emulator (black and red lines, respectively). We see that by using the linear emulator, we recover unbiased parameters at less than $0.05\sigma$. Moreover, the 1D marginalised PDFs, the parameters degeneracy, and the contours are all almost indistinguishable from those using directly \class. Overall, the results show a remarkable agreement between both approaches, which suggests that the accuracy of our linear emulator is sufficient for the analyses of forthcoming LSS surveys.

\section{Summary}\label{sec:conclusions}

In this paper, we have presented and validated a set of fast and accurate emulators aimed at speeding up the analysis of forthcoming LSS data. These emulators can provide their predictions in about one millisecond of computing time and cover a broad 8-dimensional cosmological parameter space that includes massive neutrinos and dynamical dark energy.

First, we have built and validated an emulator for the linear cold matter power spectrum, which is faster than a typical Boltzmann solver by a factor of $1000$. The accuracy of the emulator is subpercent over $10^{-4} < k < 50$ (\autoref{fig:accuracy}) and it also accurately predicts the growth of fluctuations, the baryonic acoustic oscillations, and the suppression of clustering induced by massive neutrinos (\autoref{fig:accuracy_three}). Our second set of emulators predict multiple cross-spectra of Lagrangian fields relevant for a second-order perturbative model of galaxy bias. We compute these fields with second-order Lagrangian perturbation theory which can then be emulated with percent accuracy (\autoref{fig:models_lpt}). Finally, we have shown that the accuracy of the linear emulator can be used to provide unbiased cosmological constraints for a tomographic analysis of a Euclid-like weak lensing survey (\autoref{fig:contours}).

The emulators presented in this work are part of the {\tt baccoemu} project\footnote{\url{http://bacco.dipc.org/emulator.html}.}, and have been specifically designed to be used with those we have previously built for the nonlinear matter power spectrum \citep{Angulo2020}, the modifications induced by baryonic physics \citep{Arico2020c}, and galaxy bias \citep{Zennaro2021}. All together, they can contribute towards a comprehensive, fast, and accurate cosmological exploitation of LSS data.

\section{Data and software availability} % Required
\subsection{Underlying data}
\begin{quote}
Training and validation sets of the linear matter and Lagrangian bias power spectra presented.\\
\\
\\
This project contains the following underlying data:
\begin{itemize}
	\item \href{https://bacco.dipc.org/emulator_training_sets/linear_emulator_training_set.tar}{linear\_emulator\_training\_set.tar}. 
	   Set of 240,000 linear matter power spectra obtained using the Boltzmann solver \class, and used as training and validation sets for the artificial neural network.
	\item \href{https://bacco.dipc.org/emulator_training_sets/lpt_emulator_training_set.tar}{lpt\_emulator\_training\_set.tar} Set of 20,000 Lagrangian bias power spectra, used as training and validation sets for the artificial neural network. 
\end{itemize}

Data are provided as pickle files readable in python, which contain dictionaries organised as described in the included README files. 
Data are available under the terms use of MIT license.
\end{quote}

\subsection{Software availability}
All our emulators are publicly available at \url{http://www.dipc.org/bacco}, or in the bitbucket repository \url{https://bitbucket.org/rangulo/baccoemu}, under the terms of MIT license. \\
The Boltzmann solver \class and \camb are freely available at \url{https://lesgourg.github.io/class_public/class.html}, \url{https://camb.info/}.
%If you are describing new software, please make the source code available on a Version Control System (VCS) such as GitHub, BitBucket or SourceForge, and provide details of the repository and the license under which the software can be used in the article.

%\section{Competing interests}
%No competing interests were disclosed.

%\section{Grant information}
%This project has received funding from the European Research Council (ERC) under the European Union’s Horizon 2020 research and innovation programme (grant agreement No 716151). REA acknowledges the support from the Generalitat Valenciana project of excellence Prometeo/2020/085.

\section{Acknowledgements}
We warmly thank Jon\'{a}s Chaves-Montero, Sergio Contreras, Marcos Pellejero-Iba\~nez, and Jens St\"ucker, for useful discussions.

{\small\bibliographystyle{unsrtnat}
\bibliography{bibliography}}

\begin{thebibliography}{48}
\providecommand{\natexlab}[1]{#1}
\providecommand{\url}[1]{\texttt{#1}}
\expandafter\ifx\csname urlstyle\endcsname\relax
  \providecommand{\doi}[1]{doi: #1}\else
  \providecommand{\doi}{doi: \begingroup \urlstyle{rm}\Url}\fi

\bibitem[{Seljak} and {Zaldarriaga}(1996)]{Seljak:1996}
Uros {Seljak} and Matias {Zaldarriaga}.
\newblock {A Line-of-Sight Integration Approach to Cosmic Microwave Background
  Anisotropies}.
\newblock \emph{\apj}, 469:\penalty0 437, October 1996.
\newblock \doi{10.1086/177793}.

\bibitem[{Lewis} et~al.(2000){Lewis}, {Challinor}, and {Lasenby}]{Lewis:2000}
Antony {Lewis}, Anthony {Challinor}, and Anthony {Lasenby}.
\newblock {Efficient Computation of Cosmic Microwave Background Anisotropies in
  Closed Friedmann-Robertson-Walker Models}.
\newblock \emph{\apj}, 538\penalty0 (2):\penalty0 473--476, August 2000.
\newblock \doi{10.1086/309179}.

\bibitem[{Lesgourgues}(2011{\natexlab{a}})]{Lesgourgues2011}
Julien {Lesgourgues}.
\newblock {The Cosmic Linear Anisotropy Solving System (CLASS) I: Overview}.
\newblock \emph{arXiv e-prints}, art. arXiv:1104.2932, Apr 2011{\natexlab{a}}.

\bibitem[{Blas} et~al.(2011){Blas}, {Lesgourgues}, and {Tram}]{Blas2011}
Diego {Blas}, Julien {Lesgourgues}, and Thomas {Tram}.
\newblock {The Cosmic Linear Anisotropy Solving System (CLASS). Part II:
  Approximation schemes}.
\newblock \emph{\jcap}, 2011\penalty0 (7):\penalty0 034, July 2011.
\newblock \doi{10.1088/1475-7516/2011/07/034}.

\bibitem[{Heitmann} et~al.(2014){Heitmann}, {Lawrence}, {Kwan}, {Habib}, and
  {Higdon}]{Heitmann2014}
K.~{Heitmann}, E.~{Lawrence}, J.~{Kwan}, S.~{Habib}, and D.~{Higdon}.
\newblock {The Coyote Universe Extended: Precision Emulation of the Matter
  Power Spectrum}.
\newblock \emph{\apj}, 780:\penalty0 111, January 2014.
\newblock \doi{10.1088/0004-637X/780/1/111}.

\bibitem[{McClintock} and {Rozo}(2019)]{McClintock2019b}
Thomas {McClintock} and Eduardo {Rozo}.
\newblock {Reconstructing probability distributions with Gaussian processes}.
\newblock \emph{\mnras}, 489\penalty0 (3):\penalty0 4155--4160, November 2019.
\newblock \doi{10.1093/mnras/stz2426}.

\bibitem[{Bocquet} et~al.(2020){Bocquet}, {Heitmann}, {Habib}, {Lawrence},
  {Uram}, {Frontiere}, {Pope}, and {Finkel}]{Bocquet2020}
Sebastian {Bocquet}, Katrin {Heitmann}, Salman {Habib}, Earl {Lawrence}, Thomas
  {Uram}, Nicholas {Frontiere}, Adrian {Pope}, and Hal {Finkel}.
\newblock {The Mira-Titan Universe. III. Emulation of the Halo Mass Function}.
\newblock \emph{\apj}, 901\penalty0 (1):\penalty0 5, September 2020.
\newblock \doi{10.3847/1538-4357/abac5c}.

\bibitem[{Knabenhans} et~al.(2019){Knabenhans}, {Stadel}, {Marelli}, {Potter},
  {Teyssier}, {Legrand}, {Schneider}, {Sudret}, {Blot}, {Awan}, {Burigana},
  {Carvalho}, {Kurki-Suonio}, {Sirri}, and {Euclid
  Collaboration}]{EuclidEmulator}
Mischa {Knabenhans}, Joachim {Stadel}, Stefano {Marelli}, Doug {Potter}, Romain
  {Teyssier}, Laurent {Legrand}, Aurel {Schneider}, Bruno {Sudret}, Linda
  {Blot}, Saeeda {Awan}, Carlo {Burigana}, Carla~Sofia {Carvalho}, Hannu
  {Kurki-Suonio}, Gabriele {Sirri}, and {Euclid Collaboration}.
\newblock {Euclid preparation: II. The EUCLIDEMULATOR - a tool to compute the
  cosmology dependence of the nonlinear matter power spectrum}.
\newblock \emph{\mnras}, 484\penalty0 (4):\penalty0 5509--5529, Apr 2019.
\newblock \doi{10.1093/mnras/stz197}.

\bibitem[{Euclid Collaboration} et~al.(2020){Euclid Collaboration},
  {Knabenhans}, {Stadel}, {Potter}, {Dakin}, {Hannestad}, {Tram}, {Marelli},
  {Schneider}, {Teyssier}, {Andreon}, {Auricchio}, {Baccigalupi},
  {Balaguera-Antol{\'\i}nez}, {Baldi}, {Bardelli}, {Battaglia}, {Bender},
  {Biviano}, {Bodendorf}, {Bozzo}, {Branchini}, {Brescia}, {Burigana},
  {Cabanac}, {Camera}, {Capobianco}, {Cappi}, {Carbone}, {Carretero},
  {Carvalho}, {Casas}, {Casas}, {Castellano}, {Castignani}, {Cavuoti},
  {Cledassou}, {Colodro-Conde}, {Congedo}, {Conselice}, {Conversi}, {Copin},
  {Corcione}, {Coupon}, {Courtois}, {Da Silva}, {de la Torre}, {Di Ferdinando},
  {Duncan}, {Dupac}, {Fabbian}, {Farrens}, {Ferreira}, {Finelli}, {Frailis},
  {Franceschi}, {Galeotta}, {Garilli}, {Giocoli}, {Gozaliasl},
  {Graci{\'a}-Carpio}, {Grupp}, {Guzzo}, {Holmes}, {Hormuth}, {Israel},
  {Jahnke}, {Keihanen}, {Kermiche}, {Kirkpatrick}, {Kubik}, {Kunz},
  {Kurki-Suonio}, {Ligori}, {Lilje}, {Lloro}, {Maino}, {Marggraf}, {Markovic},
  {Martinet}, {Marulli}, {Massey}, {Mauri}, {Maurogordato}, {Medinaceli},
  {Meneghetti}, {Metcalf}, {Meylan}, {Moresco}, {Morin}, {Moscardini},
  {Munari}, {Neissner}, {Niemi}, {Padilla}, {Paltani}, {Pasian}, {Patrizii},
  {Pettorino}, {Pires}, {Polenta}, {Poncet}, {Raison}, {Renzi}, {Rhodes},
  {Riccio}, {Romelli}, {Roncarelli}, {Saglia}, {S{\'a}nchez}, {Sapone},
  {Schneider}, {Scottez}, {Secroun}, {Serrano}, {Sirignano}, {Sirri}, {Stanco},
  {Sureau}, {Tallada Cresp{\'\i}}, {Taylor}, {Tenti}, {Tereno}, {Toledo-Moreo},
  {Torradeflot}, {Valenziano}, {Valiviita}, {Vassallo}, {Viel}, {Wang},
  {Welikala}, {Whittaker}, {Zacchei}, and {Zucca}]{EuclidEmulator2}
{Euclid Collaboration}, M.~{Knabenhans}, J.~{Stadel}, D.~{Potter}, J.~{Dakin},
  S.~{Hannestad}, T.~{Tram}, S.~{Marelli}, A.~{Schneider}, R.~{Teyssier},
  S.~{Andreon}, N.~{Auricchio}, C.~{Baccigalupi},
  A.~{Balaguera-Antol{\'\i}nez}, M.~{Baldi}, S.~{Bardelli}, P.~{Battaglia},
  R.~{Bender}, A.~{Biviano}, C.~{Bodendorf}, E.~{Bozzo}, E.~{Branchini},
  M.~{Brescia}, C.~{Burigana}, R.~{Cabanac}, S.~{Camera}, V.~{Capobianco},
  A.~{Cappi}, C.~{Carbone}, J.~{Carretero}, C.~S. {Carvalho}, R.~{Casas},
  S.~{Casas}, M.~{Castellano}, G.~{Castignani}, S.~{Cavuoti}, R.~{Cledassou},
  C.~{Colodro-Conde}, G.~{Congedo}, C.~J. {Conselice}, L.~{Conversi},
  Y.~{Copin}, L.~{Corcione}, J.~{Coupon}, H.~M. {Courtois}, A.~{Da Silva},
  S.~{de la Torre}, D.~{Di Ferdinando}, C.~A.~J. {Duncan}, X.~{Dupac},
  G.~{Fabbian}, S.~{Farrens}, P.~G. {Ferreira}, F.~{Finelli}, M.~{Frailis},
  E.~{Franceschi}, S.~{Galeotta}, B.~{Garilli}, C.~{Giocoli}, G.~{Gozaliasl},
  J.~{Graci{\'a}-Carpio}, F.~{Grupp}, L.~{Guzzo}, W.~{Holmes}, F.~{Hormuth},
  H.~{Israel}, K.~{Jahnke}, E.~{Keihanen}, S.~{Kermiche}, C.~C. {Kirkpatrick},
  B.~{Kubik}, M.~{Kunz}, H.~{Kurki-Suonio}, S.~{Ligori}, P.~B. {Lilje},
  I.~{Lloro}, D.~{Maino}, O.~{Marggraf}, K.~{Markovic}, N.~{Martinet},
  F.~{Marulli}, R.~{Massey}, N.~{Mauri}, S.~{Maurogordato}, E.~{Medinaceli},
  M.~{Meneghetti}, B.~{Metcalf}, G.~{Meylan}, M.~{Moresco}, B.~{Morin},
  L.~{Moscardini}, E.~{Munari}, C.~{Neissner}, S.~M. {Niemi}, C.~{Padilla},
  S.~{Paltani}, F.~{Pasian}, L.~{Patrizii}, V.~{Pettorino}, S.~{Pires},
  G.~{Polenta}, M.~{Poncet}, F.~{Raison}, A.~{Renzi}, J.~{Rhodes}, G.~{Riccio},
  E.~{Romelli}, M.~{Roncarelli}, R.~{Saglia}, A.~G. {S{\'a}nchez}, D.~{Sapone},
  P.~{Schneider}, V.~{Scottez}, A.~{Secroun}, S.~{Serrano}, C.~{Sirignano},
  G.~{Sirri}, L.~{Stanco}, F.~{Sureau}, P.~{Tallada Cresp{\'\i}}, A.~N.
  {Taylor}, M.~{Tenti}, I.~{Tereno}, R.~{Toledo-Moreo}, F.~{Torradeflot},
  L.~{Valenziano}, J.~{Valiviita}, T.~{Vassallo}, M.~{Viel}, Y.~{Wang},
  N.~{Welikala}, L.~{Whittaker}, A.~{Zacchei}, and E.~{Zucca}.
\newblock {Euclid preparation: IX. EuclidEmulator2 -- Power spectrum emulation
  with massive neutrinos and self-consistent dark energy perturbations}.
\newblock \emph{arXiv e-prints}, art. arXiv:2010.11288, October 2020.

\bibitem[{Kobayashi} et~al.(2020){Kobayashi}, {Nishimichi}, {Takada},
  {Takahashi}, and {Osato}]{Kobayashi:2020}
Yosuke {Kobayashi}, Takahiro {Nishimichi}, Masahiro {Takada}, Ryuichi
  {Takahashi}, and Ken {Osato}.
\newblock {Accurate emulator for the redshift-space power spectrum of dark
  matter halos and its application to galaxy power spectrum}.
\newblock \emph{\prd}, 102\penalty0 (6):\penalty0 063504, September 2020.
\newblock \doi{10.1103/PhysRevD.102.063504}.

\bibitem[{Aric{\`o}} et~al.(2021){Aric{\`o}}, {Angulo}, {Contreras},
  {Ondaro-Mallea}, {Pellejero-Iba{\~n}ez}, and {Zennaro}]{Arico2020c}
Giovanni {Aric{\`o}}, Raul~E. {Angulo}, Sergio {Contreras}, Lurdes
  {Ondaro-Mallea}, Marcos {Pellejero-Iba{\~n}ez}, and Matteo {Zennaro}.
\newblock {The BACCO simulation project: a baryonification emulator with neural
  networks}.
\newblock \emph{\mnras}, 506\penalty0 (3):\penalty0 4070--4082, September 2021.
\newblock \doi{10.1093/mnras/stab1911}.

\bibitem[{Zennaro} et~al.(2021){Zennaro}, {Angulo}, {Pellejero-Ib{\'a}{\~n}ez},
  {St{\"u}cker}, {Contreras}, and {Aric{\`o}}]{Zennaro2021}
Matteo {Zennaro}, Raul~E. {Angulo}, Marcos {Pellejero-Ib{\'a}{\~n}ez}, Jens
  {St{\"u}cker}, Sergio {Contreras}, and Giovanni {Aric{\`o}}.
\newblock {The BACCO simulation project: biased tracers in real space}.
\newblock \emph{arXiv e-prints}, art. arXiv:2101.12187, January 2021.

\bibitem[{Winther} et~al.(2019){Winther}, {Casas}, {Baldi}, {Koyama}, {Li},
  {Lombriser}, and {Zhao}]{Winther2019}
Hans~A. {Winther}, Santiago {Casas}, Marco {Baldi}, Kazuya {Koyama}, Baojiu
  {Li}, Lucas {Lombriser}, and Gong-Bo {Zhao}.
\newblock {Emulators for the nonlinear matter power spectrum beyond
  {\ensuremath{\Lambda}} CDM}.
\newblock \emph{\prd}, 100\penalty0 (12):\penalty0 123540, December 2019.
\newblock \doi{10.1103/PhysRevD.100.123540}.

\bibitem[{McClintock} et~al.(2019){McClintock}, {Rozo}, {Becker}, {DeRose},
  {Mao}, {McLaughlin}, {Tinker}, {Wechsler}, and {Zhai}]{McClintock2019}
Thomas {McClintock}, Eduardo {Rozo}, Matthew~R. {Becker}, Joseph {DeRose},
  Yao-Yuan {Mao}, Sean {McLaughlin}, Jeremy~L. {Tinker}, Risa~H. {Wechsler},
  and Zhongxu {Zhai}.
\newblock {The Aemulus Project. II. Emulating the Halo Mass Function}.
\newblock \emph{\apj}, 872\penalty0 (1):\penalty0 53, February 2019.
\newblock \doi{10.3847/1538-4357/aaf568}.

\bibitem[{Zhai} et~al.(2019){Zhai}, {Tinker}, {Becker}, {DeRose}, {Mao},
  {McClintock}, {McLaughlin}, {Rozo}, and {Wechsler}]{Zhai2019}
Zhongxu {Zhai}, Jeremy~L. {Tinker}, Matthew~R. {Becker}, Joseph {DeRose},
  Yao-Yuan {Mao}, Thomas {McClintock}, Sean {McLaughlin}, Eduardo {Rozo}, and
  Risa~H. {Wechsler}.
\newblock {The Aemulus Project. III. Emulation of the Galaxy Correlation
  Function}.
\newblock \emph{\apj}, 874\penalty0 (1):\penalty0 95, March 2019.
\newblock \doi{10.3847/1538-4357/ab0d7b}.

\bibitem[{Bird} et~al.(2019){Bird}, {Rogers}, {Peiris}, {Verde}, {Font-Ribera},
  and {Pontzen}]{Bird2019}
Simeon {Bird}, Keir~K. {Rogers}, Hiranya~V. {Peiris}, Licia {Verde}, Andreu
  {Font-Ribera}, and Andrew {Pontzen}.
\newblock {An emulator for the Lyman-{\ensuremath{\alpha}} forest}.
\newblock \emph{\jcap}, 2019\penalty0 (2):\penalty0 050, February 2019.
\newblock \doi{10.1088/1475-7516/2019/02/050}.

\bibitem[{Schneider} et~al.(2020){Schneider}, {Stoira}, {Refregier}, {Weiss},
  {Knabenhans}, {Stadel}, and {Teyssier}]{Schneider2020}
Aurel {Schneider}, Nicola {Stoira}, Alexandre {Refregier}, Andreas~J. {Weiss},
  Mischa {Knabenhans}, Joachim {Stadel}, and Romain {Teyssier}.
\newblock {Baryonic effects for weak lensing. Part I. Power spectrum and
  covariance matrix}.
\newblock \emph{\jcap}, 2020\penalty0 (4):\penalty0 019, April 2020.
\newblock \doi{10.1088/1475-7516/2020/04/019}.

\bibitem[{Angulo} et~al.(2021){Angulo}, {Zennaro}, {Contreras}, {Aric{\`o}},
  {Pellejero-Iba{\~n}ez}, and {St{\"u}cker}]{Angulo2020}
Raul~E. {Angulo}, Matteo {Zennaro}, Sergio {Contreras}, Giovanni {Aric{\`o}},
  Marcos {Pellejero-Iba{\~n}ez}, and Jens {St{\"u}cker}.
\newblock {The BACCO simulation project: exploiting the full power of
  large-scale structure for cosmology}.
\newblock \emph{\mnras}, 507\penalty0 (4):\penalty0 5869--5881, November 2021.
\newblock \doi{10.1093/mnras/stab2018}.

\bibitem[{Smith} et~al.(2003){Smith}, {Peacock}, {Jenkins}, {White}, {Frenk},
  {Pearce}, {Thomas}, {Efstathiou}, and {Couchman}]{Smith2003}
R.~E. {Smith}, J.~A. {Peacock}, A.~{Jenkins}, S.~D.~M. {White}, C.~S. {Frenk},
  F.~R. {Pearce}, P.~A. {Thomas}, G.~{Efstathiou}, and H.~M.~P. {Couchman}.
\newblock {Stable clustering, the halo model and non-linear cosmological power
  spectra}.
\newblock \emph{\mnras}, 341\penalty0 (4):\penalty0 1311--1332, June 2003.
\newblock \doi{10.1046/j.1365-8711.2003.06503.x}.

\bibitem[{Takahashi} et~al.(2012){Takahashi}, {Sato}, {Nishimichi}, {Taruya},
  and {Oguri}]{Takahashi2012}
R.~{Takahashi}, M.~{Sato}, T.~{Nishimichi}, A.~{Taruya}, and M.~{Oguri}.
\newblock {Revising the Halofit Model for the Nonlinear Matter Power Spectrum}.
\newblock \emph{\apj}, 761:\penalty0 152, December 2012.
\newblock \doi{10.1088/0004-637X/761/2/152}.

\bibitem[{Leclercq}(2018)]{Leclercq2018}
Florent {Leclercq}.
\newblock {Bayesian optimization for likelihood-free cosmological inference}.
\newblock \emph{\prd}, 98\penalty0 (6):\penalty0 063511, September 2018.
\newblock \doi{10.1103/PhysRevD.98.063511}.

\bibitem[{Pellejero-Iba{\~n}ez} et~al.(2020){Pellejero-Iba{\~n}ez}, {Angulo},
  {Aric{\'o}}, {Zennaro}, {Contreras}, and {St{\"u}cker}]{Pellejero2020}
Marcos {Pellejero-Iba{\~n}ez}, Raul~E. {Angulo}, Giovanni {Aric{\'o}}, Matteo
  {Zennaro}, Sergio {Contreras}, and Jens {St{\"u}cker}.
\newblock {Cosmological parameter estimation via iterative emulation of
  likelihoods}.
\newblock \emph{\mnras}, 499\penalty0 (4):\penalty0 5257--5268, October 2020.
\newblock \doi{10.1093/mnras/staa3075}.

\bibitem[{Albers} et~al.(2019){Albers}, {Fidler}, {Lesgourgues},
  {Sch{\"o}neberg}, and {Torrado}]{Albers:2019}
Jasper {Albers}, Christian {Fidler}, Julien {Lesgourgues}, Nils
  {Sch{\"o}neberg}, and Jesus {Torrado}.
\newblock {CosmicNet. Part I. Physics-driven implementation of neural networks
  within Einstein-Boltzmann Solvers}.
\newblock \emph{\jcap}, 2019\penalty0 (9):\penalty0 028, September 2019.
\newblock \doi{10.1088/1475-7516/2019/09/028}.

\bibitem[{Fendt} and {Wandelt}(2007)]{Fendt2007}
William~A. {Fendt} and Benjamin~D. {Wandelt}.
\newblock {Pico: Parameters for the Impatient Cosmologist}.
\newblock \emph{\apj}, 654\penalty0 (1):\penalty0 2--11, January 2007.
\newblock \doi{10.1086/508342}.

\bibitem[{Auld} et~al.(2007){Auld}, {Bridges}, {Hobson}, and {Gull}]{Auld:2007}
T.~{Auld}, M.~{Bridges}, M.~P. {Hobson}, and S.~F. {Gull}.
\newblock {Fast cosmological parameter estimation using neural networks}.
\newblock \emph{\mnras}, 376\penalty0 (1):\penalty0 L11--L15, March 2007.
\newblock \doi{10.1111/j.1745-3933.2006.00276.x}.

\bibitem[{Auld} et~al.(2008){Auld}, {Bridges}, and {Hobson}]{Auld:2008}
T.~{Auld}, M.~{Bridges}, and M.~P. {Hobson}.
\newblock {COSMONET: fast cosmological parameter estimation in non-flat models
  using neural networks}.
\newblock \emph{\mnras}, 387\penalty0 (4):\penalty0 1575--1582, July 2008.
\newblock \doi{10.1111/j.1365-2966.2008.13279.x}.

\bibitem[{Chevallier} and {Polarski}(2001)]{ChevallierPolarski2001}
Michel {Chevallier} and David {Polarski}.
\newblock {Accelerating Universes with Scaling Dark Matter}.
\newblock \emph{International Journal of Modern Physics D}, 10\penalty0
  (2):\penalty0 213--223, Jan 2001.
\newblock \doi{10.1142/S0218271801000822}.

\bibitem[{Linder}(2003)]{Linder2003}
Eric~V. {Linder}.
\newblock {Exploring the Expansion History of the Universe}.
\newblock \emph{\prl}, 90\penalty0 (9):\penalty0 091301, Mar 2003.
\newblock \doi{10.1103/PhysRevLett.90.091301}.

\bibitem[{Planck Collaboration} et~al.(2018){Planck Collaboration}, {Aghanim},
  {Akrami}, {Ashdown}, {Aumont}, {Baccigalupi}, {Ballardini}, {Banday},
  {Barreiro}, {Bartolo}, {Basak}, {Battye}, {Benabed}, {Bernard}, {Bersanelli},
  {Bielewicz}, {Bock}, {Bond}, {Borrill}, {Bouchet}, {Boulanger}, {Bucher},
  {Burigana}, {Butler}, {Calabrese}, {Cardoso}, {Carron}, {Challinor},
  {Chiang}, {Chluba}, {Colombo}, {Combet}, {Contreras}, {Crill}, {Cuttaia}, {de
  Bernardis}, {de Zotti}, {Delabrouille}, {Delouis}, {Di Valentino}, {Diego},
  {Dor{\'e}}, {Douspis}, {Ducout}, {Dupac}, {Dusini}, {Efstathiou}, {Elsner},
  {En{\ss}lin}, {Eriksen}, {Fantaye}, {Farhang}, {Fergusson},
  {Fernandez-Cobos}, {Finelli}, {Forastieri}, {Frailis}, {Franceschi},
  {Frolov}, {Galeotta}, {Galli}, {Ganga}, {G{\'e}nova-Santos}, {Gerbino},
  {Ghosh}, {Gonz{\'a}lez-Nuevo}, {G{\'o}rski}, {Gratton}, {Gruppuso},
  {Gudmundsson}, {Hamann}, {Handley}, {Herranz}, {Hivon}, {Huang}, {Jaffe},
  {Jones}, {Karakci}, {Keih{\"a}nen}, {Keskitalo}, {Kiiveri}, {Kim}, {Kisner},
  {Knox}, {Krachmalnicoff}, {Kunz}, {Kurki-Suonio}, {Lagache}, {Lamarre},
  {Lasenby}, {Lattanzi}, {Lawrence}, {Le Jeune}, {Lemos}, {Lesgourgues},
  {Levrier}, {Lewis}, {Liguori}, {Lilje}, {Lilley}, {Lindholm},
  {L{\'o}pez-Caniego}, {Lubin}, {Ma}, {Mac{\'{\i}}as-P{\'e}rez}, {Maggio},
  {Maino}, {Mandolesi}, {Mangilli}, {Marcos-Caballero}, {Maris}, {Martin},
  {Martinelli}, {Mart{\'{\i}}nez-Gonz{\'a}lez}, {Matarrese}, {Mauri}, {McEwen},
  {Meinhold}, {Melchiorri}, {Mennella}, {Migliaccio}, {Millea}, {Mitra},
  {Miville-Desch{\^e}nes}, {Molinari}, {Montier}, {Morgante}, {Moss}, {Natoli},
  {N{\o}rgaard-Nielsen}, {Pagano}, {Paoletti}, {Partridge}, {Patanchon},
  {Peiris}, {Perrotta}, {Pettorino}, {Piacentini}, {Polastri}, {Polenta},
  {Puget}, {Rachen}, {Reinecke}, {Remazeilles}, {Renzi}, {Rocha}, {Rosset},
  {Roudier}, {Rubi{\~n}o-Mart{\'{\i}}n}, {Ruiz-Granados}, {Salvati}, {Sandri},
  {Savelainen}, {Scott}, {Shellard}, {Sirignano}, {Sirri}, {Spencer},
  {Sunyaev}, {Suur-Uski}, {Tauber}, {Tavagnacco}, {Tenti}, {Toffolatti},
  {Tomasi}, {Trombetti}, {Valenziano}, {Valiviita}, {Van Tent}, {Vibert},
  {Vielva}, {Villa}, {Vittorio}, {Wandelt}, {Wehus}, {White}, {White},
  {Zacchei}, and {Zonca}]{Planck2018}
{Planck Collaboration}, N.~{Aghanim}, Y.~{Akrami}, M.~{Ashdown}, J.~{Aumont},
  C.~{Baccigalupi}, M.~{Ballardini}, A.~J. {Banday}, R.~B. {Barreiro},
  N.~{Bartolo}, S.~{Basak}, R.~{Battye}, K.~{Benabed}, J.-P. {Bernard},
  M.~{Bersanelli}, P.~{Bielewicz}, J.~J. {Bock}, J.~R. {Bond}, J.~{Borrill},
  F.~R. {Bouchet}, F.~{Boulanger}, M.~{Bucher}, C.~{Burigana}, R.~C. {Butler},
  E.~{Calabrese}, J.-F. {Cardoso}, J.~{Carron}, A.~{Challinor}, H.~C. {Chiang},
  J.~{Chluba}, L.~P.~L. {Colombo}, C.~{Combet}, D.~{Contreras}, B.~P. {Crill},
  F.~{Cuttaia}, P.~{de Bernardis}, G.~{de Zotti}, J.~{Delabrouille}, J.-M.
  {Delouis}, E.~{Di Valentino}, J.~M. {Diego}, O.~{Dor{\'e}}, M.~{Douspis},
  A.~{Ducout}, X.~{Dupac}, S.~{Dusini}, G.~{Efstathiou}, F.~{Elsner}, T.~A.
  {En{\ss}lin}, H.~K. {Eriksen}, Y.~{Fantaye}, M.~{Farhang}, J.~{Fergusson},
  R.~{Fernandez-Cobos}, F.~{Finelli}, F.~{Forastieri}, M.~{Frailis},
  E.~{Franceschi}, A.~{Frolov}, S.~{Galeotta}, S.~{Galli}, K.~{Ganga}, R.~T.
  {G{\'e}nova-Santos}, M.~{Gerbino}, T.~{Ghosh}, J.~{Gonz{\'a}lez-Nuevo}, K.~M.
  {G{\'o}rski}, S.~{Gratton}, A.~{Gruppuso}, J.~E. {Gudmundsson}, J.~{Hamann},
  W.~{Handley}, D.~{Herranz}, E.~{Hivon}, Z.~{Huang}, A.~H. {Jaffe}, W.~C.
  {Jones}, A.~{Karakci}, E.~{Keih{\"a}nen}, R.~{Keskitalo}, K.~{Kiiveri},
  J.~{Kim}, T.~S. {Kisner}, L.~{Knox}, N.~{Krachmalnicoff}, M.~{Kunz},
  H.~{Kurki-Suonio}, G.~{Lagache}, J.-M. {Lamarre}, A.~{Lasenby},
  M.~{Lattanzi}, C.~R. {Lawrence}, M.~{Le Jeune}, P.~{Lemos}, J.~{Lesgourgues},
  F.~{Levrier}, A.~{Lewis}, M.~{Liguori}, P.~B. {Lilje}, M.~{Lilley},
  V.~{Lindholm}, M.~{L{\'o}pez-Caniego}, P.~M. {Lubin}, Y.-Z. {Ma}, J.~F.
  {Mac{\'{\i}}as-P{\'e}rez}, G.~{Maggio}, D.~{Maino}, N.~{Mandolesi},
  A.~{Mangilli}, A.~{Marcos-Caballero}, M.~{Maris}, P.~G. {Martin},
  M.~{Martinelli}, E.~{Mart{\'{\i}}nez-Gonz{\'a}lez}, S.~{Matarrese},
  N.~{Mauri}, J.~D. {McEwen}, P.~R. {Meinhold}, A.~{Melchiorri}, A.~{Mennella},
  M.~{Migliaccio}, M.~{Millea}, S.~{Mitra}, M.-A. {Miville-Desch{\^e}nes},
  D.~{Molinari}, L.~{Montier}, G.~{Morgante}, A.~{Moss}, P.~{Natoli}, H.~U.
  {N{\o}rgaard-Nielsen}, L.~{Pagano}, D.~{Paoletti}, B.~{Partridge},
  G.~{Patanchon}, H.~V. {Peiris}, F.~{Perrotta}, V.~{Pettorino},
  F.~{Piacentini}, L.~{Polastri}, G.~{Polenta}, J.-L. {Puget}, J.~P. {Rachen},
  M.~{Reinecke}, M.~{Remazeilles}, A.~{Renzi}, G.~{Rocha}, C.~{Rosset},
  G.~{Roudier}, J.~A. {Rubi{\~n}o-Mart{\'{\i}}n}, B.~{Ruiz-Granados},
  L.~{Salvati}, M.~{Sandri}, M.~{Savelainen}, D.~{Scott}, E.~P.~S. {Shellard},
  C.~{Sirignano}, G.~{Sirri}, L.~D. {Spencer}, R.~{Sunyaev}, A.-S. {Suur-Uski},
  J.~A. {Tauber}, D.~{Tavagnacco}, M.~{Tenti}, L.~{Toffolatti}, M.~{Tomasi},
  T.~{Trombetti}, L.~{Valenziano}, J.~{Valiviita}, B.~{Van Tent}, L.~{Vibert},
  P.~{Vielva}, F.~{Villa}, N.~{Vittorio}, B.~D. {Wandelt}, I.~K. {Wehus},
  M.~{White}, S.~D.~M. {White}, A.~{Zacchei}, and A.~{Zonca}.
\newblock {Planck 2018 results. VI. Cosmological parameters}.
\newblock \emph{arXiv e-prints}, July 2018.

\bibitem[{Castorina} et~al.(2015){Castorina}, {Carbone}, {Bel}, {Sefusatti},
  and {Dolag}]{Castorina:2015}
Emanuele {Castorina}, Carmelita {Carbone}, Julien {Bel}, Emiliano {Sefusatti},
  and Klaus {Dolag}.
\newblock {DEMNUni: the clustering of large-scale structures in the presence of
  massive neutrinos}.
\newblock \emph{\jcap}, 2015\penalty0 (7):\penalty0 043, July 2015.
\newblock \doi{10.1088/1475-7516/2015/07/043}.

\bibitem[{Zennaro} et~al.(2019){Zennaro}, {Angulo}, {Aric{\`o}}, {Contreras},
  and {Pellejero-Ib{\'a}{\~n}ez}]{Zennaro:2019}
Matteo {Zennaro}, Ra{\'u}l~E. {Angulo}, Giovanni {Aric{\`o}}, Sergio
  {Contreras}, and Marcos {Pellejero-Ib{\'a}{\~n}ez}.
\newblock {How to add massive neutrinos to your {\ensuremath{\Lambda}}CDM
  simulation - extending cosmology rescaling algorithms}.
\newblock \emph{\mnras}, 489\penalty0 (4):\penalty0 5938--5951, November 2019.
\newblock \doi{10.1093/mnras/stz2612}.

\bibitem[{Eisenstein} and {Hu}(1999)]{EisensteinHu1999}
Daniel~J. {Eisenstein} and Wayne {Hu}.
\newblock {Power Spectra for Cold Dark Matter and Its Variants}.
\newblock \emph{\apj}, 511\penalty0 (1):\penalty0 5--15, January 1999.
\newblock \doi{10.1086/306640}.

\bibitem[Chollet et~al.(2015)]{chollet2015keras}
Francois Chollet et~al.
\newblock Keras, 2015.
\newblock URL \url{https://github.com/fchollet/keras}.

\bibitem[Abadi et~al.(2016)Abadi, Barham, Chen, Chen, Davis, Dean, Devin,
  Ghemawat, Irving, Isard, et~al.]{abadi2016tensorflow}
Mart{\'\i}n Abadi, Paul Barham, Jianmin Chen, Zhifeng Chen, Andy Davis, Jeffrey
  Dean, Matthieu Devin, Sanjay Ghemawat, Geoffrey Irving, Michael Isard, et~al.
\newblock Tensorflow: A system for large-scale machine learning.
\newblock In \emph{12th $\{$USENIX$\}$ Symposium on Operating Systems Design
  and Implementation ($\{$OSDI$\}$ 16)}, pages 265--283, 2016.

\bibitem[{Schneider} et~al.(2016){Schneider}, {Teyssier}, {Potter}, {Stadel},
  {Onions}, {Reed}, {Smith}, {Springel}, {Pearce}, and
  {Scoccimarro}]{Schneider2016}
A.~{Schneider}, R.~{Teyssier}, D.~{Potter}, J.~{Stadel}, J.~{Onions}, D.~S.
  {Reed}, R.~E. {Smith}, V.~{Springel}, F.~R. {Pearce}, and R.~{Scoccimarro}.
\newblock {Matter power spectrum and the challenge of percent accuracy}.
\newblock \emph{\jcap}, 4:\penalty0 047, April 2016.
\newblock \doi{10.1088/1475-7516/2016/04/047}.

\bibitem[{Springel} et~al.(2020){Springel}, {Pakmor}, {Zier}, and
  {Reinecke}]{Springel:2020}
Volker {Springel}, R{\"u}diger {Pakmor}, Oliver {Zier}, and Martin {Reinecke}.
\newblock {Simulating cosmic structure formation with the GADGET-4 code}.
\newblock \emph{arXiv e-prints}, art. arXiv:2010.03567, October 2020.

\bibitem[{Baumann} et~al.(2018){Baumann}, {Green}, and
  {Wallisch}]{Baumann:2018}
Daniel {Baumann}, Daniel {Green}, and Benjamin {Wallisch}.
\newblock {Searching for light relics with large-scale structure}.
\newblock \emph{\jcap}, 2018\penalty0 (8):\penalty0 029, August 2018.
\newblock \doi{10.1088/1475-7516/2018/08/029}.

\bibitem[{Matsubara}(2008)]{Matsubara2008}
Takahiko {Matsubara}.
\newblock {Nonlinear perturbation theory with halo bias and redshift-space
  distortions via the Lagrangian picture}.
\newblock \emph{\prd}, 78\penalty0 (8):\penalty0 083519, October 2008.
\newblock \doi{10.1103/PhysRevD.78.083519}.

\bibitem[{Modi} et~al.(2020){Modi}, {Chen}, and {White}]{Modi2020}
Chirag {Modi}, Shi-Fan {Chen}, and Martin {White}.
\newblock {Simulations and symmetries}.
\newblock \emph{\mnras}, 492\penalty0 (4):\penalty0 5754--5763, March 2020.
\newblock \doi{10.1093/mnras/staa251}.

\bibitem[{Kokron} et~al.(2021){Kokron}, {DeRose}, {Chen}, {White}, and
  {Wechsler}]{Kokron2021}
Nickolas {Kokron}, Joseph {DeRose}, Shi-Fan {Chen}, Martin {White}, and Risa~H.
  {Wechsler}.
\newblock {The cosmology dependence of galaxy clustering and lensing from a
  hybrid $N$-body-perturbation theory model}.
\newblock \emph{arXiv e-prints}, art. arXiv:2101.11014, January 2021.

\bibitem[{Chen} et~al.(2020){Chen}, {Vlah}, and {White}]{Chen2020}
Shi-Fan {Chen}, Zvonimir {Vlah}, and Martin {White}.
\newblock {Consistent modeling of velocity statistics and redshift-space
  distortions in one-loop perturbation theory}.
\newblock \emph{\jcap}, 2020\penalty0 (7):\penalty0 062, July 2020.
\newblock \doi{10.1088/1475-7516/2020/07/062}.

\bibitem[{Amendola} et~al.(2018){Amendola}, {Appleby}, {Avgoustidis}, {Bacon},
  {Baker}, {Baldi}, {Bartolo}, {Blanchard}, {Bonvin}, {Borgani}, {Branchini},
  {Burrage}, {Camera}, {Carbone}, {Casarini}, {Cropper}, {de Rham}, {Dietrich},
  {Di Porto}, {Durrer}, {Ealet}, {Ferreira}, {Finelli},
  {Garc{\'{\i}}a-Bellido}, {Giannantonio}, {Guzzo}, {Heavens}, {Heisenberg},
  {Heymans}, {Hoekstra}, {Hollenstein}, {Holmes}, {Hwang}, {Jahnke},
  {Kitching}, {Koivisto}, {Kunz}, {La Vacca}, {Linder}, {March}, {Marra},
  {Martins}, {Majerotto}, {Markovic}, {Marsh}, {Marulli}, {Massey}, {Mellier},
  {Montanari}, {Mota}, {Nunes}, {Percival}, {Pettorino}, {Porciani},
  {Quercellini}, {Read}, {Rinaldi}, {Sapone}, {Sawicki}, {Scaramella},
  {Skordis}, {Simpson}, {Taylor}, {Thomas}, {Trotta}, {Verde}, {Vernizzi},
  {Vollmer}, {Wang}, {Weller}, and {Zlosnik}]{Amendola2018}
L.~{Amendola}, S.~{Appleby}, A.~{Avgoustidis}, D.~{Bacon}, T.~{Baker},
  M.~{Baldi}, N.~{Bartolo}, A.~{Blanchard}, C.~{Bonvin}, S.~{Borgani},
  E.~{Branchini}, C.~{Burrage}, S.~{Camera}, C.~{Carbone}, L.~{Casarini},
  M.~{Cropper}, C.~{de Rham}, J.~P. {Dietrich}, C.~{Di Porto}, R.~{Durrer},
  A.~{Ealet}, P.~G. {Ferreira}, F.~{Finelli}, J.~{Garc{\'{\i}}a-Bellido},
  T.~{Giannantonio}, L.~{Guzzo}, A.~{Heavens}, L.~{Heisenberg}, C.~{Heymans},
  H.~{Hoekstra}, L.~{Hollenstein}, R.~{Holmes}, Z.~{Hwang}, K.~{Jahnke}, T.~D.
  {Kitching}, T.~{Koivisto}, M.~{Kunz}, G.~{La Vacca}, E.~{Linder}, M.~{March},
  V.~{Marra}, C.~{Martins}, E.~{Majerotto}, D.~{Markovic}, D.~{Marsh},
  F.~{Marulli}, R.~{Massey}, Y.~{Mellier}, F.~{Montanari}, D.~F. {Mota}, N.~J.
  {Nunes}, W.~{Percival}, V.~{Pettorino}, C.~{Porciani}, C.~{Quercellini},
  J.~{Read}, M.~{Rinaldi}, D.~{Sapone}, I.~{Sawicki}, R.~{Scaramella},
  C.~{Skordis}, F.~{Simpson}, A.~{Taylor}, S.~{Thomas}, R.~{Trotta},
  L.~{Verde}, F.~{Vernizzi}, A.~{Vollmer}, Y.~{Wang}, J.~{Weller}, and
  T.~{Zlosnik}.
\newblock {Cosmology and fundamental physics with the Euclid satellite}.
\newblock \emph{Living Reviews in Relativity}, 21:\penalty0 2, April 2018.
\newblock \doi{10.1007/s41114-017-0010-3}.

\bibitem[{Barreira} et~al.(2018){Barreira}, {Krause}, and
  {Schmidt}]{Barreira2018}
Alexandre {Barreira}, Elisabeth {Krause}, and Fabian {Schmidt}.
\newblock {Accurate cosmic shear errors: do we need ensembles of simulations?}
\newblock \emph{\jcap}, 2018\penalty0 (10):\penalty0 053, October 2018.
\newblock \doi{10.1088/1475-7516/2018/10/053}.

\bibitem[{Foreman-Mackey} et~al.(2013){Foreman-Mackey}, {Hogg}, {Lang}, and
  {Goodman}]{Foreman2013}
Daniel {Foreman-Mackey}, David~W. {Hogg}, Dustin {Lang}, and Jonathan
  {Goodman}.
\newblock {emcee: The MCMC Hammer}.
\newblock \emph{\pasp}, 125\penalty0 (925):\penalty0 306, Mar 2013.
\newblock \doi{10.1086/670067}.

\bibitem[Lewis and Bridle(2002)]{Lewis2002}
Antony Lewis and Sarah Bridle.
\newblock {Cosmological parameters from CMB and other data: A Monte Carlo
  approach}.
\newblock \emph{\prd}, 66:\penalty0 103511, 2002.
\newblock \doi{10.1103/PhysRevD.66.103511}.

\bibitem[{Lesgourgues}(2011{\natexlab{b}})]{classIII}
Julien {Lesgourgues}.
\newblock {The Cosmic Linear Anisotropy Solving System (CLASS) III: Comparision
  with CAMB for LambdaCDM}.
\newblock \emph{arXiv e-prints}, art. arXiv:1104.2934, April
  2011{\natexlab{b}}.

\bibitem[{Lesgourgues} and {Tram}(2011)]{classIV}
Julien {Lesgourgues} and Thomas {Tram}.
\newblock {The Cosmic Linear Anisotropy Solving System (CLASS) IV: efficient
  implementation of non-cold relics}.
\newblock \emph{\jcap}, 2011\penalty0 (9):\penalty0 032, September 2011.
\newblock \doi{10.1088/1475-7516/2011/09/032}.

\bibitem[{Zennaro} et~al.(2017){Zennaro}, {Bel}, {Villaescusa-Navarro},
  {Carbone}, {Sefusatti}, and {Guzzo}]{Zennaro2017}
M.~{Zennaro}, J.~{Bel}, F.~{Villaescusa-Navarro}, C.~{Carbone}, E.~{Sefusatti},
  and L.~{Guzzo}.
\newblock {Initial conditions for accurate N-body simulations of massive
  neutrino cosmologies}.
\newblock \emph{\mnras}, 466\penalty0 (3):\penalty0 3244--3258, Apr 2017.
\newblock \doi{10.1093/mnras/stw3340}.

\end{thebibliography}

\appendix

\section{Appendix: Class setup}\label{app:setup}

\begin{figure*}
\begin{center}
\includegraphics[width=.9\textwidth]{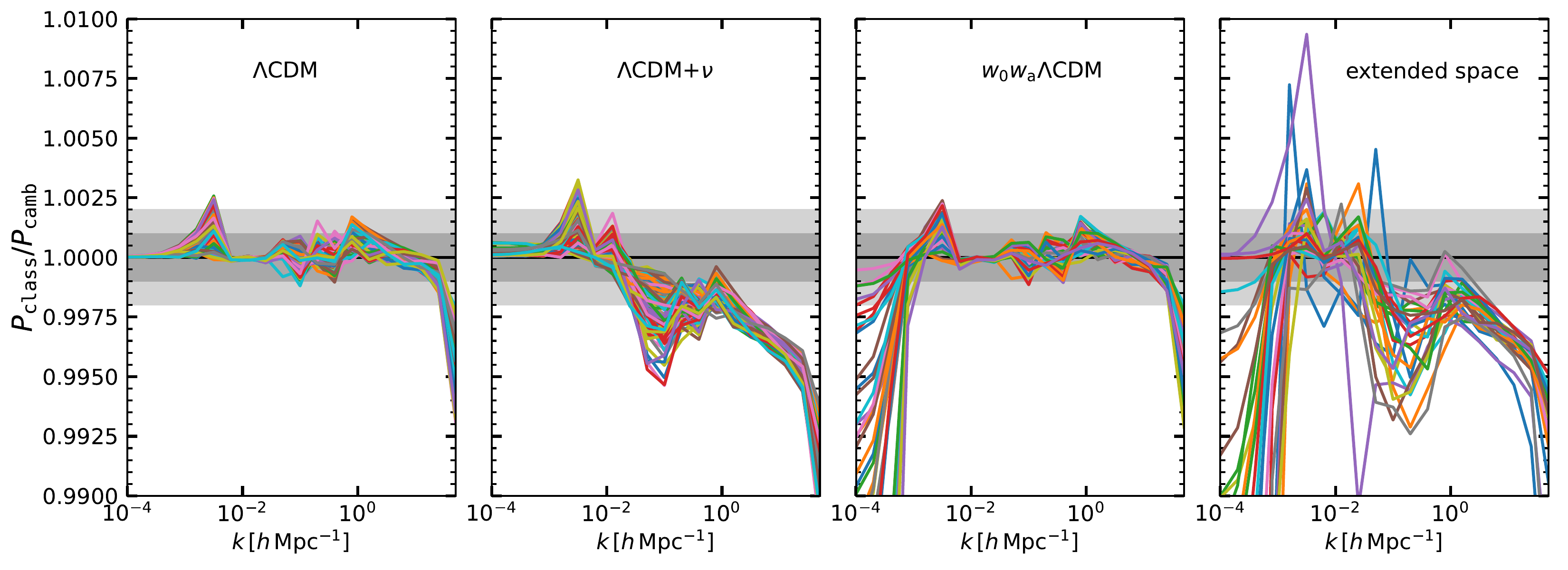}
\caption{Comparison of the linear power spectrum provided by \class and \camb, two independent Bolztmann solvers. Each panel displays the ratio for multiple cosmologies in the minimal $\Lambda$CDM model, $\Lambda$CDM  plus neutrinos, $\Lambda$CDM  plus dynamical dark energy, and in $\Lambda$CDM plus neutrinos and dynamical dark energy. Shaded regions denote 0.1 and 0.02\% differences.
}
\label{fig:class_vs_camb}
\end{center}
\end{figure*}

In this Appendix we provide some details on our \class setup and compare its predictions against those of \camb.

In our \class calculations, we set the primordial helium fraction to $Y_{\rm He}=0.24$, the optical depth at reionisation
$\tau_{\rm reio}=0.0952$, the number of (degenerate) massive neutrinos to $N_\nu=3$, and the number of relativistic species $N_r=3.046$. The neutrino to photon temperature is computed accounting for non-instantaneous neutrino decoupling and spectral distortions induced by the reheating, $T_{\nu}=0.71611$. We assume the standard \class neutrino fluid approximation, and solve the neutrino integrals with an automatic choice of the best quadrature method, employing 150 momentum bins.

To get an estimate of the absolute accuracy of the \class power spectra, we make a comparison against the predictions of another Boltzmann solver, \camb \citep{Lewis2002}. We use in \camb, whenever possible, an analogous setup, specifying a degenerate neutrino hierarchy, and furthermore setting to True the {\it accurate\_massive\_neutrino\_transfer} and {\it Reionization} options.

We compare both codes in 100 points distributed as a Latin hyper-cube, in a pure $\Lambda$CDM scenario, adding massive neutrinos, in a $\Lambda$CDM plus $w_0 w_{\rm a}$ dynamical dark energy, and finally in our full {\it extended} cosmological space. In \autoref{fig:class_vs_camb} we show the results of this comparison.

In the minimal $\Lambda$CDM scenario and in the $w_0 w_{\rm a}$ cosmologies, the two codes agree at the $0.1\%$ level, except for $k > 10\ihMpc$ where \camb underestimates the \class solution by 0.2\%, likely due to \camb neglecting the impact of reionisation on the baryon sound speed \citep{classIII}.
When considering the massive neutrinos, the agreement degrades to a $0.5\%$, a value larger than what found in \cite{classIV} but in agreement with \cite{Zennaro2017}, and approaches $1\%$ at the smallest scaled employed. When dropping the \class fluid approximation, the agreement between the two codes improves to 0.2\% at $k < 10^{-1} \ihMpc$, but get increasingly worse at smaller scales, reaching 1\% at scales $k \approx 40 \ihMpc$. 
Within the dynamical dark energy models, the two codes agree mostly at $0.1\%$, except for scales $k < 10^{-3} \ihMpc$ and $k > 30\ihMpc$, where the difference can reach $1\%$. 
Although the scales probed by the next generation LSS surveys are well recovered by both the Boltzmann solvers considered here, a more in-depth investigation of the differences between \class and \camb within a large cosmological space and wide range of scales is necessary, since we find percent differences that might potentially impact future analyses. We leave such investigation for future works.

\end{document}